\documentstyle[preprint,aps,eqsecnum,epsfig]{revtex}
\begin{document}
\draft
\title {Muon-Spin Rotation Spectra in the Mixed Phase of
High-T$_c$ Superconductors : Thermal Fluctuations and 
Disorder Effects}
\author{Gautam I. Menon\cite{byimsc}}
\address{Department of Physics,\\
Simon Fraser University, Burnaby\\
B.C. V5A 1S6, Canada}
\author{Chandan Dasgupta\cite{byjnc} and T. V.
Ramakrishnan\cite{byjnc}} 
\address{Department of Physics, Indian Institute of Science,\\
Bangalore -- 560 012, India.}
\date{\today}
\maketitle
\begin{abstract}
We study muon-spin rotation ($\mu$SR) spectra in the mixed phase
of highly anisotropic layered superconductors, specifically
$Bi_{2+x}Sr_{2-x}CaCu_2O_{8+\delta}$ (BSCCO), by modeling the
fluid and solid phases of pancake vortices using liquid-state
and density functional methods.  The role of thermal
fluctuations in causing motional narrowing of $\mu$SR lineshapes
is quantified in terms of a first-principles theory of the
flux-lattice melting transition.  The effects of random point
pinning are investigated using a replica treatment of liquid
state correlations and a replicated density functional theory.
Our results indicate that motional narrowing in the pure system,
although substantial, cannot account for the remarkably small
linewidths obtained experimentally at relatively high fields and
low temperatures. We find that satisfactory agreement with the
$\mu$SR data for BSCCO in this regime can be obtained through
the {\it ansatz} that this ``phase'' is characterized by {\it
frozen} short-range positional correlations reflecting the
structure of the liquid just above the melting transition. This
proposal is consistent with recent suggestions of a ``pinned
liquid'' or ``glassy'' state of pancake vortices in the presence
of pinning disorder. Our results for the high-temperature liquid phase
indicate that measurable linewidths may be obtained in this phase
as a consequence of density inhomogeneities induced by the pinning
disorder. The results presented here comprise a unified,
first-principles theoretical treatment of $\mu$SR spectra in
highly anisotropic layered superconductors in terms of a
controlled set of approximations.
\end{abstract}
\section{Introduction}
\label{sec1}

In the mixed phase of a type-II superconductor, an externally
applied magnetic field penetrates the bulk of the sample in the
form of lines of magnetic flux \cite{abr}.  The distribution function 
$n(B)\/$ of the local magnetic induction
$B$ in the sample is thus a non-trivial quantity, depending on the flux
distribution associated with a single vortex line, as well as on the
arrangement of the vortex lines in space and time.  This
distribution function can be obtained through muon-spin rotation
($\mu$SR) experiments based on the following procedure: An
ensemble of muons, initially polarized transverse to the
magnetic induction, is implanted uniformly throughout the sample
\cite{schenck,brandt2,soviet}. Since the local magnetic
induction varies in space, the spins of the muons in different regions of the
sample precess at different rates, resulting in a dephasing of
the initial polarization with time. This dephasing is monitored
by tracking the angular distribution of the positrons emitted as
the muons decay. The measured polarization can then be related, via a
Fourier transform, to $n(B)$.

The utility of the $\mu$SR method is that it provides an
accurate local probe of magnetic field variations in the bulk
irrespective of whether such variations are periodic (as for a
regular vortex lattice) or have more complex structure
reflecting arrangements of vortex lines with varying degrees of
spatial correlations. In high-temperature superconductors such
as $Bi_{2+x}Sr_{2-x}CaCu_2O_{8+\delta}$ (BSCCO), $\mu$SR 
experiments carried out in a field perpendicular to the superconducting
layers (this is the geometry we consider throughout this paper) 
have provided indirect evidence for
what is believed to be a melting transition \cite{rev} of the Abrikosov
flux-line lattice to a flux liquid with short-ranged positional
correlations. In the experiments of Lee {\it et al}
\cite{lee,lee1,lee2} on BSCCO, where $n(B)\/$ is obtained as
a function of the external field $H\/$ and the temperature
$T\/$, the location of the melting line at fields of the order
of 10 mT is inferred from the abrupt changes in the $\mu$SR
linewidths and lineshapes which occur as $H$ and/or $T$ are
varied across the transition boundary. In this low-field regime,
the results obtained are in broad agreement with extensive
experimental and theoretical work over the past decade aimed at
detecting and understanding the nature of such a melting
transition. In contrast, the nature of the transition and of the
low-temperature phase at {\it high} fields still remains
controversial.

For a regular Abrikosov lattice at zero temperature and in the
absence of disorder, the second moment of $n(B)\/$ can be simply
related to the magnetic penetration depth $\lambda$
\cite{brandt2}. For a wide variety of superconductors, the values of
$\lambda\/$ obtained in this way \cite{sonier1,riseman,sonier2}
agree closely with those obtained by other means \cite{keller}.
However, for BSCCO, such an interpretation of the high field
data yields values for the $ab$ plane
penetration depth $\lambda_{ab}$ ranging between 3500 and 4500
$\AA$ \cite{harsh1}. These numbers are to be contrasted to
typical values obtained by other means which suggest 
$1100 < \lambda_{ab} < 2800\AA\/$, with values
in the range $1500 < \lambda_{ab} < 2000 \AA$ believed to be
appropriate for slightly overdoped samples 
\cite{lee1,lee2,schneider,bernhard,aegerter}.
This discrepancy originates in the anomalously small values of
the linewidth $\sqrt{[\Delta B^2]}$ (the brackets $[\cdots]$ denote a
space average) obtained in experiments
performed at high fields \cite{data}. Fits to the lineshapes at
low fields, in contrast, yield results in reasonable agreement
with conventionally accepted values for $\lambda_{ab}$ with
$\lambda_{ab} \simeq 1800\AA$ obtained in the recent experiments
of Lee {\it et al.} \cite{lee,lee1,lee2}. The magnitude of the
linewidths is found to change rather sharply across a
``critical'' value $H_{cr}$ ($\sim $ 100 mT) of the magnetic field
which depends only weakly on
the temperature (the ``low'' and ``high'' field regimes referred
to earlier can be more precisely correlated to $H_{cr}$). As the
external field is increased across $H_{cr}$, the $\mu$SR lineshapes
are observed to become far more
symmetric, concomitant with the sharp reduction in linewidths.
This asymmetry can be quantified through the normalized
``asymmetry parameter'' $\alpha = [\Delta B^3]^{1/3}/[\Delta
B^2]^{1/2}$. In the high field regime, measured values of
$\alpha$ range from 0.5 -- 0.9 ($\alpha \simeq 1.2$ in the
Abrikosov flux lattice phase). These observations suggest that
the nature of the flux array changes qualitatively as the
applied field is increased across $H_{cr}$.  The existence of
such a critical field value across which the attributes of the
flux-line system change abruptly has been confirmed in a variety
of other measurements, including transport \cite{trans} and
magnetization \cite{magn} experiments and neutron scattering \cite{neutr}.
Possible explanations for these observations include the
proposal of a sharp crossover from three-dimensional to effectively
two-dimensional behavior as a function of the
field in the absence of disorder \cite{glazman}, an entanglement
transition of the flux-line lattice \cite{ertaz,nels} as the
field is increased, and the transition between a pinning-induced,
weakly disordered ``Bragg glass'' phase \cite{braggl} at 
low field values\cite{giamarchi,kierfield} to 
a strongly disordered ``pinned liquid'' or ``vortex glass'' \cite{f2h}
phase at high fields. Figs.~\ref{fig1}--\ref{fig3} 
illustrate schematic phase diagrams.
The phase diagram of a pure system according to 
the conventional Abrikosov theory is shown in Fig.~\ref{fig1}. A phase
diagram illustrating the Abrikosov lattice and  vortex liquid 
phases expected for a
pure system is shown in Fig.~\ref{fig2}, and Fig.~\ref{fig3} shows
a conjectured phase diagram in the presence of random point pinning,
illustrating Bragg glass, pinned liquid (or vortex glass) and
vortex liquid phases.

The explanations proposed for reduced linewidths in the high
field regime of BSCCO fall into two major classes which are not
completely unrelated.  One class of theories suggests that
reduced linewidths arise due to thermal broadening (and possibly quantum
broadening -- we ignore quantum effects in the bulk of this paper,
but discuss them briefly in the concluding section) of density 
distributions in the lattice phase. It is argued that the effects 
of thermal fluctuations in BSCCO 
are enhanced above $H_{cr}$ due to the nearly
two-dimensional character of vortex fluctuations in this
regime -- the flux-line structure in BSCCO is more accurately 
thought of in terms of ``pancake'' vortices \cite{clem} moving on
the superconducting layers and interacting via a combination
of electromagnetic and Josephson couplings.
Such behavior should generalize to
all sufficiently anisotropic, layered compounds \cite{brandt3}
and should be reflected to a lesser extent in more isotropic materials.
Calculations representing this class of theories 
have so far been restricted to harmonic treatments of elastic
fluctuations about the zero-temperature Abrikosov flux 
lattice \cite{lee1,lee2,brandt3,harsh2}. 
In these calculations, the effects of thermal fluctuations in
the lattice phase are expressed in terms of the
Lindemann parameter $L = \sqrt{\langle u^2 \rangle}/a_0$, where
$\langle u^2 \rangle$ is the mean square thermal fluctuation of a flux-line
from its equilibrium position at melting and $a_0$ is the
mean inter-line spacing. In the analysis of experimental data,
$L$ is treated as a phenomenological fitting parameter. 
One purpose of this paper is to present calculations of $\mu$SR 
lineshapes and linewidths based on a detailed and microscopic 
theory of flux-liquid and flux-lattice phases in a highly anisotropic
layered superconductor. This theory can be used to {\em calculate} the 
Lindemann parameter as well as density distributions at
freezing. Such density distributions feed into a
calculation of the lineshapes measurable in $\mu$SR
experiments -- results of a calculation of these lineshapes are 
presented here.

The second class of theories invokes the presence of underlying quenched
disorder and proposes random pinning as a possible source of reduced 
linewidths. The quasi-two-dimensional vortex system
above $H_{cr}$ is argued to be highly susceptible to pinning. Several
proposals for the nature of this low-temperature state
exist in the literature. The ``vortex glass'' description \cite{f2h}
concentrates on the physics at large length scales
and thus does not address issues of short-range order. There has
been much recent interest in the proposal \cite{giamarchi} that
at low fields, the flux-line system is in a ``Bragg glass''
phase in which the $\delta$-function Bragg peaks associated with the pure
crystalline phase are replaced by algebraic singularities. 
Such a relatively ordered phase is expected
\cite{ertaz,giamarchi,kierfield} to yield 
to a topologically disordered phase (where 
dislocations proliferate) at fields larger
than a critical value, this transition being driven 
by the increasing relevance of disorder due to the effective reduction of the
dimensionality at high fields. Evidence for such a transition
has been found in numerical simulations \cite{ryu} of a system
of flux lines in the presence of random point pinning. 
The Bragg glass scenario is a plausible explanation of 
the experimental data. However, the precise nature of the phase obtained
at large fields remains unclear. 
Given that the physics of the problem at high fields 
is dominated by the fluctuations of pancake vortices weakly
coupled across layers,  correlations of vortices across the 
superconducting layers are expected to be much weaker
than correlations between vortices on the same layer.
Specifically in the context of $\mu$SR experiments
on BSCCO, some analytic work exists on a model which 
assumes that the low-temperature state in the presence of pinning
disorder is characterized by a disordered
stacking of ordered planes \cite{harsh2,oldmodel}. However, 
it is difficult to see how even relatively strong disorder could
stabilize such an arrangement in which the misalignment
of ordered layers is penalized by a macroscopic (proportional
to the area of the layer for each layer) energy cost. It is 
reasonable to expect that some fraction of this cost is relaxed by allowing
dislocations to form on each layer. If
a non-zero density of dislocations is allowed to form, the detailed 
microscopic arrangement of vortices must be qualitatively
different from that proposed in such simplified models. 
In this paper we propose that a reasonable assumption for this
disordered state is that it resembles a ``frozen'' liquid in the
sense that the positional correlations in this state are essentially the
same as that in a liquid just above its putative freezing
transition, but the time scale for structural relaxation is much
longer than that in the liquid.

In our theoretical analysis of
$\mu$SR spectra in the mixed phase of extremely anisotropic,
layered, type-II superconductors in a field applied perpendicular to
the superconducting layers, the flux-solid phase is described using the
density functional theory \cite{ry,ysingh,us1,us2} and the fluid
phase using a liquid state-theory adapted by us \cite{us1,us2}
to study ``pancake'' vortices in the Lawrence-Doniach
description \cite{ldon} of a layered superconductor in the limit of
vanishing Josephson coupling. Our
approach assumes a value of $\lambda_{ab}$
consistent with the values obtained from low-field $\mu$SR
and other experiments on BSCCO 
($1500 \AA < \lambda_{ab} < 1800\AA$) and calculates 
linewidths and lineshapes appropriate to these values,
incorporating thermal fluctuations, disorder effects (we
consider random point pinning produced by atomic-scale pinning centers
such as oxygen defects \cite{chudnovsky}) as well as
corrections due to the finite core size \cite{yaouanc}. 
The approach outlined
here should be applicable to the calculation of
$\mu$SR lineshapes in artificial
superconducting multilayers \cite{multi} in which the
distance between neighboring superconducting layers 
can be tuned so as to reduce
the Josephson coupling between layers to an insignificant level. 
The approach of this paper is, in principle, non-perturbative
in that our description of thermally broadened density
distributions in the flux-solid phase is not restricted to
assuming a harmonic form for the free energy cost associated 
with small deviations from the perfect lattice.

Our principal results for the pure system include a calculation 
of $\mu$SR lineshapes and
linewidths at freezing. These results can be extended below
the freezing transition by {\it assuming} harmonic fluctuations
at sufficiently low-temperatures, whose magnitude is consistent with the
theoretically calculated $\langle u^2 \rangle$ at the melting transition.
We find that thermal broadening should reduce the 
$\mu$SR linewidth for BSCCO at the flux-lattice melting temperature 
by a factor of about 4 relative to that for a perfect Abrikosov lattice. 
These results quantify thermal fluctuation amplitudes in the pure system and 
the magnitude of the corresponding corrections
to the linewidths. These corrections can then be used to
estimate $\lambda_{ab}$ more accurately. However, the linewidths
so obtained are still far in excess of those measured experimentally for 
$H \gg H_{cr}$ even at the ``melting'' transition.
Further, theories based on thermal fluctuations about an ordered
triangular lattice of flux lines cannot explain the anomalously
small linewidths obtained at low temperatures, as pointed out
by several other authors. Thus, disorder effects appear to be
necessary for an explanation of the $\mu$SR linewidths at high fields.

We also study, using a replicated liquid-state and density
functional theory \cite{disprl}, the effects of weak pinning
disorder on the $\mu$SR linewidth in both the high-temperature liquid and
the low-temperature solid. We find that such disorder should not affect
$\mu$SR lineshapes considerably in the low-temperature state
{\it if it has strong local crystalline order}, such as might be
expected in a state composed
of ``Larkin domains''\cite{larkin} of characteristic size much larger than
the average inter-vortex spacing, and in a Bragg glass
phase with algebraic decay of positional correlations. The fact that
pinning should affect the lineshapes only weakly in such a state
is expected on intuitive grounds because only the
{\it short-range} order of the arrangement of the vortices is 
probed in $\mu$SR experiments. However, these results
disagree with experimental observations in the high-field limit,
indicating that the system cannot have strong short-range crystalline
order in this regime. In the absence of a detailed theory of the
structure of the flux-line system in this regime, we propose that
{\it the positional correlations in this state resemble
those in the liquid just above the solid-liquid ``transition''}. This state,
however, differs from the high-temperature liquid in that the
characteristic time scale for structural relaxation is much longer than
that in the liquid. The correlations probed in the $\mu$SR  experiments
are then naturally understood as being representative of the positional
correlations in the equilibrated liquid. This proposal is consistent
with the ``pinned liquid'' state postulated in the
phase diagram suggested in Ref.\cite{giamarchi}. This description should
also be appropriate if the high-field, low-temperature state of the system
represents a new glassy phase that is separated from the
high-temperature liquid by a line of true phase transitions. This is because
the positional correlations in a structural glass are generally quite
similar to those in the liquid near the glass transition.
We present calculations of the second and third moments
(the third moment is calculated using a specific decoupling 
approximation for correlations) of the field
distribution in such a low-temperature disordered
state and find that the results (summarized in Table I) are in 
good agreement with the experimental data on BSCCO. 

Several time scales are relevant to the analysis of 
transverse $\mu$SR spectra in the mixed phase. 
A muon injected into the sample at time
$t=0\/$ thermalizes due to interactions with the bulk material
on timescales of order $10^{-10}~s\/$\cite{brandt2}. This
time scale is much
shorter than the muon lifetime and typical precession times
($\sim 10^{-6} - 10^{-7}~s$). After thermalization, the muon diffuses
in the sample. If the diffusivity of the muons is small, the typical
distance traveled by a muon during its lifetime
is much smaller than both the mean inter-vortex
spacing and the penetration depth. In this limit, which is the
one we consider, the polarization after time $t\/$ is completely 
determined by the initial polarization and the distribution of
local fields. (A complication that may be present in real experiments
is that the muons may be preferentially absorbed at certain sites. 
We do not consider this possibility in the present work 
and assume that the muons are uniformly distributed.)
Other time scales that enter the problem are the relaxation times
of density fluctuations in the solid and liquid phases, 
which determine the time scales over which the field felt by a
stationary muon varies. For the pure system, consistent with other
work and with results obtained on YBa$_2$Cu$_3$O$_{7-\delta}$~(YBCO)
and other high-T$_c$ superconductors,
we assume that the relaxation time in the low-temperature (lattice)
phase is very short relative to the inverse precession rate of 
the muons. (Experimentally, these time scales are
found to be of order $10^{-10}s$\cite{song} close to the
flux-lattice melting transition.) In the fluid phase, 
the $\mu$SR linewidth would be expected to be vanishingly
small due to motional narrowing if the vortex lines are highly mobile.
However, if the vortex lines have very low mobility,
such as might be expected in strongly pinned, entangled or glassy
states, the second moment of $n(B)$ can be related to
the static structure factor of the liquid. We argue that
the low mobility situation is realized at low temperatures
in the high-field state of the pancake vortex system in the
presence of random point pinning. At high temperatures and deep 
into the fluid regime, 
we argue that the vortex lines are highly mobile, so that the 
structural relaxation times of the vortex system are much shorter than
the inverse precession rate of the muons. Even in this regime,
there is a non-vanishing contribution to linewidths that comes from
a space-averaged correlation function of pinning-induced 
density inhomogeneities in the liquid. We present a calculation of this 
contribution to the second moment of the field distribution function. Our
results are representative of the linewidths obtained 
experimentally deep into the liquid phase.

The outline of this paper is as follows. In section \ref{sec2}, we briefly
discuss the calculation of the field distribution and
its moments in a perfect Abrikosov lattice.
In section \ref{sec3}, we introduce our model of ``pancake'' vortices 
confined to the superconducting planes, and briefly describe 
the liquid-state and density functional methods
used in our study. We also discuss the replica method we use to
study the disordered liquid and its freezing transition in the
presence of random point pinning.
In section \ref{sec4}, we calculate the magnetic field
distribution and its moments for a thermally broadened Abrikosov 
lattice. We use the density functional description of the 
crystalline solid phase of the vortex
system \cite{us1,us2} in our analysis. In section \ref{sec5},
we calculate these moments in a ``frozen'' liquid state of the vortex
system. We provide calculations of the second and third
moments to support our proposal that
values of $[\Delta B^2]$ consistent with $\lambda_{ab}$
values of about $1800\AA$ can be obtained if it is
assumed that the low-temperature state of the vortex
system at high fields is a glassy one characterized by short-ranged,
liquid-like correlations and long relaxation times. 
The effects of weak disorder on
the second moment of the field distribution in the high-temperature
liquid state are discussed in section \ref{sec6}, where we point
out that non-vanishing linewidths can
be obtained in this state by considering the contribution from the static 
inhomogeneities induced in the liquid by the quenched
disorder. Section \ref{sec7} contains a summary of the main results obtained
in this paper and a few concluding remarks.

\section{$\mu$SR Spectrum for an Ideal Abrikosov Lattice}
\label{sec2}

The time evolution of the transverse polarization $(P^+({\bf r},t)
= P_x({\bf r},t) + iP_y({\bf r},t)\/$) of a
muon in a locally inhomogenous magnetic field ${\bf B}({\bf
r})\/$, assumed to lie in the $\hat{z}$ direction 
(i.e. ${\bf B}({\bf r})=\hat{z}B({\bf r})\/$), is
given by the Bloch-Torrey equation \cite{torrey}
\begin{equation}
\frac{\partial P^+({\bf r},t)}{\partial t}=-i\gamma_\mu B({\bf
r})P^+({\bf r},t)+D\nabla^2 P^+({\bf r},t),
\label{eq21}
\end{equation}
where $D\/$ is the muon diffusivity, $P_x({\bf r},t),P_y({\bf r},t)\/$ 
are the two components of the polarization transverse to the field and
$\gamma_\mu\/$ is the muon gyromagnetic ratio (=
8.516$\times 10^8$~rad~T$^{-1}$~s$^{-1})\/$. Eq.(\ref{eq21}) is to be
solved subject to the initial condition 
$P^+({\bf r},t=0)=P^+_i({\bf r})\/$. If the initial distribution
of muons is uniform then $P^+({\bf
r},t=0)=P^0\/$ and if, in addition, the muon does not diffuse, the
solution of Eq.(\ref{eq21}) is
\begin{equation}
P^+({\bf r},t)=P^0\exp[-i\gamma_\mu B({\bf r}) t].
\label{eq22}
\end{equation}
The space averaged distribution of the local magnetic induction 
is defined as
\begin{equation}
n(B) = [\delta(B-B({\bf r})],
\label{eq23}
\end{equation}
where $[A({\bf r})]$ refers to
the space average $(1/V)\int_V d{\bf r}A({\bf r})\/$, $V$ being the
volume of the sample. It then follows that
\begin{equation}
P^{(+)}(t)=\frac{P^{(0)}}{\gamma_\mu}\int^{+\infty}_{-\infty}
n\left(\frac{\omega} {\gamma_\mu}\right)\exp(-i\omega t)d\omega,
\label{eq24}
\end{equation}
where 
$P^{(+)}(t)\/$ is the spatial average of $P^{(+)}({\bf r},t)$ and
$\gamma_\mu B = \omega$. The measured $P^{(+)}(t)\/$ 
can then be used to obtain, through an inverse Fourier 
transformation, the distribution of internal fields in the sample.

The local magnetic field at a point ${\bf r}\/$ due to an
array of straight vortex lines in the $\hat z$ direction is given by
\begin{equation}
B({\bf r})=\int d{\bf r}^\prime~b(\mid{\bf r}-{\bf
r}^\prime\mid) \rho({\bf r}^\prime),
\label{eq25}
\end{equation}
where $b({\bf r}-{\bf r}^\prime)\/$ denotes the field at
point {\bf r} due to a single vortex line at ${\bf r}^\prime$
and $\rho({\bf r})\/$ is the flux line density at point
${\bf r}$ ({\bf r} is a two-dimensional vector perpendicular to
the $z$-axis).  Using the Fourier decomposition $\rho({\bf
r}^\prime)= (1/A)\sum_{\bf k} \rho_{\bf k}\exp(i{\bf k \cdot
r^\prime})\/$, and $b({\bf r})= (1/A)\sum_{\bf k} b_{\bf
k}\exp(i{\bf k\cdot r})\/$ (A is the cross sectional area of the
sample) we get 
\begin{equation} 
B({\bf r})=(1/A)\sum_{\bf k} b_{\bf k}\rho_{\bf k}\exp(i{\bf k
\cdot r}). 
\label{eq26}
\end{equation}
In an Abrikosov lattice, only terms involving reciprocal
lattice vectors ({\bf k} ={\bf G}) survive in Eq.(\ref{eq26}). For
a perfect lattice with $\delta$-function densities, 
$\rho_{\bf G}=N\/$ for all ${\bf G}\/$ ($N\/$
is the number of flux lines) and the
sum may be written as 
\begin{equation} 
B({\bf r})=\frac{B_0}{\Phi_0}\sum_{\bf G} b_{\bf G}\exp(i{\bf G
\cdot r}), 
\label{eq27}
\end{equation}
where $N\Phi_0=B_0A\/$ and $B_0$ is the magnetic induction 
within the sample,
averaged over distances of order the penetration depth. $\Phi_0$ is the
quantum unit of flux carried by each line and equals $2.07 \times $10$^{-15}$ 
T~m$^2$. In the London model,
\begin{equation}
b_{\bf G} = \frac{\Phi_0}{1 + \lambda^2 G^2},
\label{eq28}
\end{equation}
with $G^2 = G_x^2 + G_y^2$,
which leads to a magnetic induction that diverges logarithmically
at the core of a vortex. This unphysical divergence can be
eliminated by introducing a core ``form factor'' $f(G)$ which represents
the behavior near the core more accurately. This then gives
\begin{equation}
b_{\bf G} = \frac{\Phi_0 f(G)}{1 + \lambda^2 G^2},
\label{eq29}
\end{equation}
Analytic expressions for $f(G)$ can be obtained from solutions
of the full Ginzburg-Landau equations at low inductions. A
variational solution to these equations by Clem \cite{clem2} yields the
result $f(G) = g K_1(g)$, where $g = \sqrt{2}\xi(G^2 +
\lambda^{-2})^{1/2}$. Here, $\xi$ is the coherence length of the
superconductor and $K_1(x)$ is a modified Bessel function. A less accurate
but more commonly used representation of this core form
factor replaces $gK_1(g)$ by the Gaussian form $\exp(-\xi^2G^2/2)$.
Recently, Yaouanc {\it et al} \cite{yaouanc} 
have argued that
it is essential to retain the full form of the core form
factor for a satisfactory treatment. Using results
originally derived by Hao {\it et al} \cite{hao}, they obtain
\begin{equation}
B({\bf r})=B_0\sum_{\bf G}\frac{w(G)K_1(w(G))(1-b^4)}{(1+\lambda^2
G^2)}\exp(i{\bf G \cdot r}), 
\label{eq210}
\end{equation}
where $w^2(G) = 2(\xi^2G^2)(1+b^4)[1 -2b(1-b)^2]$, 
$b=B_0/B_{c2}$, and we have
specialized their expressions to the case in which there
are no in-plane anisotropies.
For the triangular Abrikosov lattice, the sum in
Eq.(\ref{eq210}) is to be carried out over the reciprocal lattice
generated by the vectors 
${\bf G}_1=(4\pi/\sqrt{3}a_0)\hat{y}\/$ and
${\bf G}_2=(2\pi/a_0)[\hat{x}+\hat{y}/\sqrt{3}]\/$. Here  the inter-vortex
spacing $a_0\/$ is given by $a_0^2 = 2\Phi_0/\sqrt{3}B_0\/$. For
the calculations reported in this paper, we use the full
form factor proposed by Yaouanc {\it et al.} in our
description of lineshapes and linewidths in both
solid and fluid phases.  Also, since
we work at inductions much less than $B_{c2}$, we set
$b^4 \simeq 0$ and $[1-2b(1-b)^2] \simeq 1$.

The moments of the field distribution function $n(B)\/$ can be
obtained from Eq.(\ref{eq210}). The second moment of the
distribution function $[\Delta B^2] \equiv
\int^{+\infty}_{-\infty}n(B)(B-B_0)^2dB$ is obtained as 
\begin{equation}
[\Delta B^2]=B^2_0\sum_{{\bf G}\neq 0}\frac{w^2(G)K^2_1(w(G))}
{(1+\lambda^2G^2)^2}
\simeq B^2_0\sum_{{\bf G}\neq
0}\frac{w^2(G)K^2_1(w(G))}{\lambda^4G^4}, 
\label{eq211}
\end{equation}
where 
\begin{equation}
G^2\/= \frac{16 \pi^2}{3a_0^2}[m^2+mn+n^2],
\label{eq212}
\end{equation}
$m,n\/$ are integers and we have used $\lambda/a_0 \gg 1\/$
in obtaining the final answer in Eq.(\ref{eq211}). The third moment,
defined by $[\Delta B^3] \equiv
\int^{+\infty}_{-\infty}n(B)(B-B_0)^3dB$ is obtained as 
\begin{equation}
[\Delta B^3]=B^3_0\sum_{\stackrel{{\bf G}_1,{\bf G}_2,{\bf G}_3 \neq
0} {{\bf G}_1+{\bf G}_2+{\bf G}_3 =
0}}\frac{\sigma(G_1)\sigma(G_2)\sigma(G_3)}
{(1+\lambda^2G^2_1)(1+\lambda^2G^2_2)(1+\lambda^2G^2_3)},
\label{eq213}
\end{equation}
where the $\sigma(G_i)$, $i$ = 1, 2, 3, represent the cutoff corrections
and are given by
\begin{equation}
\sigma(G_i) = w(G_i)K_1(w(G_i)).
\label{eq214}
\end{equation}

\section{Model and methods of calculation} 
\label{sec3}

In extremely anisotropic,
layered superconductors in their type-II regime in an external field
applied perpendicular to the layer plane, the
statistical mechanics of the interacting flux-line system
reduces to the problem of interacting planar
vortices. These ``pancake''
vortices are formed at the intersections of the flux lines with
the planes on which the superconducting order parameter 
$\Psi(r)\/$ attains its maximum value. In
the limit of infinite anisotropy (a reasonably accurate starting point
for BSCCO in which the anisotropy factor
$\gamma$ has been estimated to be higher than 100), 
the pancake vortices interact via 
a pair-wise potential that can be calculated within the
framework of a London model\cite{feigel1}. This limit corresponds
to assuming that the direct Josephson coupling between two adjacent 
superconducting layers is vanishingly small.
However, vortices on different layers continue
to be coupled through their electromagnetic interaction. 
Recent studies \cite{lee2,aegerter} have shown that in BSCCO, the 
electromagnetic interaction between vortices
dominates over the interaction generated by the
Josephson coupling if the temperature is lower than about 0.8$T_c(0)$. 
As the low-temperature behavior will be our primary
concern in this paper, our assumption
of infinite anisotropy can be expected to be quite 
reasonable for BSCCO.

As we have argued earlier
\cite{us1,us2}, the statistical mechanics of the flux-line system
in its fluid phase is most easily approached in this limit by 
considering the system of pancake vortices to be a classical, 
anisotropic liquid. The
equilibrium density correlation functions of this classical liquid may
be calculated by generalizing methods of liquid-state theory, and the
freezing transition of this liquid may be studied using well
known methods of classical density functional theory 
with the calculated liquid-state correlation functions as
input. Such an approach makes it possible to characterize thermally
broadened density distributions in the equilibrium solid at the
freezing transition without using any {\it ad hoc} assumption.

In this section, we first define the effective ``Hamiltonian''
of the vortex system in the
limit of infinite anisotropy, and then provide brief
introductions to the liquid-state and density functional
methods used in our analysis. 
Finally, we discuss the techniques we use to study the effects
of disorder on liquid state correlations and the extension
of density functional methods to deal with the problem of
freezing in the presence of random pinning.

\subsection{Interaction between pancake vortices in the limit of infinite
anisotropy}
If the coherence length of a layered superconductor in the
direction perpendicular to the layers is of the order of the inter-layer
spacing $d$, the appropriate generalization of the
Ginzburg-Landau functional is a free-energy functional suggested by
Lawrence and Doniach \cite{ldon}. In the limit of infinite
anisotropy (i.e. vanishing Josephson coupling between the
superconducting layers), the Lawrence-Doniach functional in the
``phase-only'' (London) approximation reduces to a particularly
simple form from which the interaction energy of a collection of
pancake vortices can be readily derived \cite{feigel1,bul}. The 
energy turns out to be a sum of pairwise interactions (this
simplification occurs only in the limit of infinite anisotropy
-- the effective ``Hamiltonian'' of a system of pancake vortices can not be
broken up into a sum of pairwise terms if the Josephson coupling
is finite). In Fourier space, the inter-vortex interaction is given by
\begin{eqnarray} 
\beta V({\bf k}) = \frac{\Gamma
\lambda^2[k_{\perp}^2+(4/d^2)\sin^2(k_zd/2)]}{k_{\perp}^2[1+\lambda^2
k_{\perp}^2+4(\lambda^2/d^2) \sin^2(k_z d/2)]},
\label{eq31}
\end{eqnarray}
with $\Gamma = \beta d \Phi^2_0/4 \pi \lambda^2\/$ ($\lambda$ is
the in-plane penetration depth -- we have removed the subscript
$ab$ for ease of notation) and $\beta=1/k_B T\/$. In
Eq.(\ref{eq31}), $k_\perp\/$ and $k_z\/$ refer, as before, to
directions in the plane of the layers and normal to them.
Two vortices lying on the same layer repel each other with a potential 
that is a logarithmic function of their separation, whereas two vortices on
different layers {\em attract} each other with a potential that is weaker than
the intralayer potential by the factor $d/\lambda$. The interlayer potential
is logarithmic in the in-plane separation and falls off exponentially in the
$z$-direction as $e^{-nd/\lambda}$ where $n$ is the layer separation.

The Fourier transforms of the radial and $z-$axis components of
the magnetic field due to a single pancake vortex at the origin are given by
\begin{equation}
{\bf b}_\perp=-\frac{d\Phi_0k_z {\bf
k}_\perp}{k^2_\perp(1+\lambda^2k^2_\perp+\lambda^2k^2_z)},
\label{eq32}
\end{equation}
and
\begin{equation}
b_z=\frac{d\Phi_0}{(1+\lambda^2k^2_\perp+\lambda^2k^2_z)},
\label{eq33}
\end{equation}
The calculation of local field distributions and their moments in
this case proceeds in much the same way as outlined in the
previous section. In this description, however, there is also a
radial component associated with the total magnetic induction.
(This component vanishes in the case of perfectly aligned straight
vortex lines). In addition, the $z$-axis
component of the magnetic induction is a weak function of $z$,
even for a straight line of pancake vortices. If the inter-layer
spacing is small, however, in comparison to the length scale
over which the field varies (i.e. $d \ll \lambda\/$), this
variation can be neglected.  

\subsection{Liquid state correlations and density functional
theory for the pure vortex system}

A basic quantity that appears in a statistical description of
the equilibrium properties of a simple isotropic liquid is
the pair distribution function $g(\mid{\bf r}-{\bf r}^\prime\mid)\/$,
defined by $\langle \textstyle \sum_{i=1}^N \sum_{j \neq i}^N
\delta({\bf r}-{\bf r}_i)\delta({\bf r}^\prime-{\bf
r}_j)/\rho^2_\ell\rangle\/$. Here, ${\bf r}_i$ denotes the location of
the $i$th particle in the liquid and $\rho_\ell$ is the average density.
The pair correlation function $h(r)$ defined by
$h(r)=g(r)-1\/$, and the static structure 
factor defined by $S(k)=1+\rho_\ell\int d{\bf r} h(r)
\exp(i{\bf k \cdot r})$ are closely related to $g(r)$. 
The pair correlation function can be decomposed as
\begin{eqnarray}
h(r) = C(r)+\rho_\ell \int d {\bf r}^{\prime} 
C(\mid {\bf r}-{\bf r}^\prime\mid)h({\bf r}^\prime),
\label{eq34}
\end{eqnarray}
which is the Ornstein-Zernike relation \cite{hanmac}. 
The Fourier transform of the direct pair correlation function $C(r)\/$ 
is related to the structure factor through
$S(k)=1/(1-\rho_\ell C(k))$. 

These definitions can be readily generalized \cite{us1,us2} to describe the
equilibrium correlations in a layered liquid of pancake
vortices. Since this system is highly anisotropic, the two-point
correlation functions in position space have two arguments: the in-plane
separation $\rho$ and the separation $nd$ in the $z$-direction
($n$ is an integer). The Fourier transforms of these functions
depend separately on $k_z$ and $k_\perp$. As discussed in detail
in Refs. \cite{us1} and \cite{us2}, standard methods of liquid
state theory, such as the Hypernetted Chain approximation (HNC) 
\cite{hanmac} and the Rogers-Young closure scheme \cite{roy}, can
be generalized to calculate these correlation functions of
the layered vortex liquid. These correlation functions are then
used as input to a density functional theory of freezing from which 
the freezing temperature and the 
time-averaged density distributions in the solid phase at
the freezing transition are obtained. 

In the density functional theory, the free energy of an inhomogeneous
configuration of the time-averaged density field
is expressed in terms of fluid-phase correlation
functions. Freezing into a crystalline structure occurs when the 
free energy of the state with the appropriate periodic density modulation
equals the free energy of the uniform liquid. 
In the density functional theory of Ramakrishnan and
Yussouff\cite{ry}, the grand-canonical free energy
cost (with respect to the uniform liquid) of producing a 
time-averaged density inhomogeneity is expressed as a 
functional of the density $\rho({\bf r})\/$.
The simplest such functional for an isotropic system is 
\begin{equation}
\frac{\Delta \Omega}{k_B T}=\int d{\bf r}
\left[\rho({\bf r})\ln \frac{\rho({\bf
r})}{\rho_\ell} - \delta \rho({\bf
r})\right] 
-\frac{1}{2}\int d{\bf r} \int d{\bf r}^\prime C(\mid
{\bf r-r^\prime} \mid) \delta \rho({\bf r}) \delta \rho({\bf
r^\prime}) + \ldots ,
\label{eq35}
\end{equation}
where $\rho({\bf r})\/$ is the density at point ${\bf r}$, 
$\delta \rho({\bf r}) \equiv \rho({\bf r}) - \rho_\ell\/$,
$T$ is the temperature, $\rho_\ell\/$ is the liquid 
density and the ellipsis denotes higher order terms 
which are conventionally set to zero.  

In mean-field theory, density configurations which
represent the equilibrium phase minimize Eq.(\ref{eq35}),
i.e. satisfy $\delta\Delta\Omega/\delta\rho({\bf r})= 0$.
This condition gives
\begin{eqnarray}
\ln \biggl[\frac{\rho({\bf r})}{\rho_\ell} \biggr] = 
\int d{\bf r}^\prime C(\mid{\bf
r-r^\prime}\mid)[\rho({\bf r^\prime})- \rho_\ell].
\label{eq36}
\end{eqnarray}
A periodic (crystalline) density configuration is a
solution of the mean-field equations if
the following self-consistency condition is satisfied:
\begin{eqnarray}
1+\eta+{\displaystyle\sum_{\bf G \neq 0}}\rho_{\bf
G} \exp(i {\bf G \cdot r}) = \exp \left[{\tilde C}_0\eta+
{\displaystyle\sum_{\bf G \neq 0}} {\tilde C}_{\bf G} 
\rho_{\bf G} \exp(i {\bf G} \cdot {\bf r})\right],
\label{eq37}
\end{eqnarray}
where the {\bf G}s are the reciprocal lattice vectors of the
structure to which the liquid freezes, $\rho_{\bf G}\/$s are
the Fourier components of the density field with wave vector
{\bf G} and we have defined ${\tilde C} \equiv \rho_\ell C\/$. 
The quantity $\eta\/$ is the fractional volume change
on freezing from the liquid. Note that the uniform liquid, for which 
$\rho_{\bf G} = 0\/$
for all {\bf G} $\ne 0$ and $\eta = 0$, is always a local minimum of the
free energy. However, because of the 
non-linearity of the self-consistency
condition, periodic density-wave solutions which could be
absolute minima of the free energy of Eq.(\ref{eq35}) may appear as the
correlations in the liquid increase. 
The $\rho_{\bf G}$s which minimize the
density functional represent the (mean-field) density
distribution in the crystalline solid. Since the
density distribution near a lattice point 
is not restricted to be Gaussian in a fully
self-consistent theory, the thermal broadening of the peaks of the
density distribution includes at least some of the anharmonic
effects that are expected to be present near the melting transition. Our
treatment, therefore, goes beyond the harmonic approximation made in several
existing studies \cite{lee1,lee2,brandt3,harsh2} 
of $\mu$SR spectrum in the mixed phase.

The density functional formalism outlined above can be
generalized in a straightforward manner \cite{us1,us2} to a
layered system of pancake vortices. The reciprocal lattice
vectors that characterize the Abrikosov lattice are $({\bf
G}_\perp,G_z=0)$ where the ${\bf G}_\perp$s are {\it two
dimensional} reciprocal lattice vectors defining a triangular
lattice. 

\subsection{Replica liquid state theory and density functional
theory in the presence of pinning disorder} 

In earlier work \cite{disprl}, two of us developed a replica theory of
the correlations of a vortex liquid in the presence of random
point pinning and derived a replicated
free-energy functional analogous to the Ramakrishnan-Yussouff
functional to describe the freezing of this disordered
liquid. We briefly summarize our methods here. Our analysis is based on
the replica method \cite{nld} applied to a system of classical particles
interacting via the Hamiltonian
\begin{equation}
H = H_{kinetic} + \frac{1}{2}\sum_{i \ne j} V(\mid{\bf r}_i - {\bf r}_j\mid)
+ \sum V_d({\bf r}_i),
\label{eq38}
\end{equation}
where $V(r)$ is a two-body interaction
potential between the particles and $V_d({\bf r})$ is
a quenched, random, one-body potential. We assume that $V_d({\bf r_i})$
is drawn from a Gaussian distribution of zero mean and short
ranged correlations: $[V_d({\bf r})V_d({\bf r^\prime})] =
K(\mid{\bf r} - {\bf r}^\prime\mid)$, with $[\cdots]$ denoting
an average over the disorder.
Using $[\ln Z] = \lim_{n \rightarrow 0} [(Z^n -1)/n]$, one obtains,
prior to taking the $n \rightarrow 0$ limit, a replicated
and disorder averaged partition function of the form
\begin{equation}
Z^R = \frac{1}{(N!)^n}\int\Pi d{\bf r}_i^\alpha
\exp(-\frac{1}{2k_BT}\sum_{\alpha=1}^n\sum_{\beta=1}^n
\sum_{i = 1}^N\sum_{j=1}^N
V^{\alpha\beta}(\mid{\bf r}_i^\alpha - {\bf r}^\beta_j\mid)).
\label{eq39}
\end{equation}
Here $\alpha,\beta$ are replica indices  and
$V^{\alpha\beta}(\mid{\bf r}^\alpha_i - {\bf r}^\beta_j\mid)
=V(\mid{\bf r}^\alpha_i - {\bf r}^\beta_j\mid)\delta_{\alpha\beta} 
-\beta K(\mid{\bf r}^\alpha_i -{\bf r}^\beta_j\mid)$.

Our approach to the problem defined by Eq.(\ref{eq39}) begins by
recognizing that it resembles the partition function
of a system of $n$ ``species'' of particles, each labeled by an appropriate
replica index. These particles interact via a
two-body interaction which depends both on particle coordinates
$({\bf r}_i,{\bf r}_j)$
and replica indices $(\alpha,\beta)$. 
This system of $n$ species of particles can be treated in
liquid state theory by considering it to be a $n$-component
mixture and taking the $n \rightarrow 0$ limit in the
Ornstein-Zernike equations governing the properties of the
mixture. In the $n \rightarrow 0$ limit, assuming
replica symmetry, we get the equations
\begin{equation}
h^{(1)}(k) = \frac{C^{(1)}(k) - \rho_\ell [C^{1)}(k) - C^{(2)}(k)]^2}{[1 -
\rho_\ell C^{(1)}(k) + \rho_\ell C^{(2)}(k)]^2},
\label{eq310}
\end{equation}
and
\begin{equation}
h^{(2)}(k) = \frac{C^{(2)}(k)}{[1 - \rho_\ell C^{(1)}(k)
+ \rho_\ell C^{(2)}(k)]^2}.
\label{eq311}
\end{equation}
These equations are written in terms of the pair correlation
functions $h^{\alpha\beta}$ and the direct correlation
functions $C^{\alpha\beta}$ of the replicated system.
The assumption of replica symmetry implies that
$C^{\alpha\beta} = C^{(1)}\delta_{\alpha\beta} + C^{(2)}(1 -
\delta_{\alpha\beta})$ and
$h^{\alpha\beta} = h^{(1)}\delta_{\alpha\beta} + h^{(2)}(1 -
\delta_{\alpha\beta})$. The physical interpretation of these
correlation functions is as follows. The function $h^{(1)}$ describes
the disorder-averaged equal-time (equilibrium) correlation of
fluctuations of the local density, and $h^{(2)}$ represents the 
disorder-averaged correlation of disorder-induced deviations of the 
time-averaged local density from its average value $\rho_\ell$.

In Ref.\cite{disprl}, we calculated the functions $h^{(1)}(\rho,nd)$
and $h^{(2)}(\rho,nd)$ for the layered vortex system using the
HNC closure approximation. To estimate the inter-replica
interaction $\beta K(\rho,nd)$, we assumed that the principal
source of disorder is atomic scale pinning centers such as
oxygen defects \cite{chudnovsky} 
which act to reduce $T_c$ locally. This yields
\begin{equation}
\beta K(\rho,nd) \simeq
\Gamma^\prime\exp(-\rho^2/\xi^2)\delta_{n,0}. 
\label{eq318}
\end{equation}
Here $\xi \simeq 15 \AA$ is the coherence length in the $ab$-plane, and 
$\Gamma^\prime \approx 10^{-5}\Gamma^2$ for point pinning of
strength $dr_0^2H_c^2/8\pi$, where $r_0$ is an atomic scale length
parameter and we have assumed defect densities of the order of
$10^{20}$/cm$^3$. In Figs.~\ref{fig4} and \ref{fig5}, we show 
typical examples of the calculated 
correlation functions $g^{(1)}$ and $g^{(2)}$ obtained using 
this theory at two different values of $T$ at fixed
induction $B_0$.

The calculated correlation functions $C^{(1)}(r)$ and
$C^{(2)}(r)$ can then be used as input into an appropriately
generalized version \cite{ysingh} of the density functional 
theory for a mixture of $n$ species of particles. This leads to the
following density functional in the $n\rightarrow 0$ limit:
\begin{eqnarray}
\frac{\Delta \Omega}{k_B T}&=&\int d{\bf r}
\left[\rho({\bf r})\ln \frac{\rho({\bf
r})}{\rho_\ell} - \delta \rho({\bf
r})\right] \nonumber \\
&&-\frac{1}{2}\int d{\bf r} \int d{\bf r}^\prime [C^{(1)}(\mid
{\bf r}-{\bf r}^\prime \mid) - C^{(2)}({\bf r}-{\bf r}^\prime \mid)]
[\rho({\bf r})-\rho_\ell][\rho({\bf
r^\prime})-\rho_\ell] \ldots.
\label{eq312}
\end{eqnarray}
Here we have assumed that the density field is the same in all the
replicas ($\rho^\alpha({\bf r}) = \rho({\bf r})$ for all $\alpha$). 

Generalization of this free energy to a layered system of pancake
vortices is straightforward. In Ref.\cite{disprl}, 
we {\em assumed} that if the pinning disorder is
weak, then the ``solid'' phase of the vortex system is described by a
global minimum of this free energy characterized by non-zero
values of the crystalline order parameters $\rho_{\bf G}$. Since we are
interested in a mean-field description of a first-order transition (the
flux-lattice melting transition is experimentally found to be first
order if the pinning disorder is weak), this assumption is justified as
long as the low-temperature state exhibits strong local crystalline
order (specifically, as long as the translational correlation length in
the low-temperature phase is much longer than the range of positional
correlations in the liquid just above the freezing transition). 
Results of a full calculation of the phase-boundary obtained through 
this method were presented in Ref.\cite{disprl}. Our
principal result was that the liquid-solid phase boundary in the
$(B-T)$ plane in the absence of disorder is only mildly affected by 
the presence of weak disorder, the suppression of this phase boundary
to lower temperatures being larger at higher fields. These predictions
have been confirmed in recent experimental studies \cite{khy} of this
phase boundary in electron-irradiated samples of BSCCO. In the
present work, we use the results obtained in Ref.\cite{disprl} for
the density distribution in the ``nearly crystalline'' phase to 
analyze the effects of weak pinning disorder on $\mu$SR lineshapes 
in the low-temperature, low-field state of the vortex system.

\section{$\mu$SR spectrum in the crystalline phase of BSCCO}
\label{sec4}

In this section, we first consider the crystalline (Abrikosov lattice) phase
of the vortex system in the absence of pinning disorder.
If the typical time scales for phonon-like density fluctuations in this
state are much shorter than the characteristic time
scale of precession  of the muon spin ($\sim (\gamma_\mu B)^{-1}\/$), the
muons see a broadened (equilibrium) density distribution of flux lines.
The time-averaged local magnetic field $B({\bf r})\/$ is then obtained 
through the convolution relation
\begin{equation}
B({\bf r})=\int d{\bf r}^\prime 
b(\mid{\bf r}-{\bf r}^\prime\mid)\bar{\rho}({\bf r^\prime}),
\label{eq41}
\end{equation}
which differs from Eq.(\ref{eq25}) in that 
$\bar{\rho}({\bf r})\/$ now represents the {\it time-averaged}
density distribution. The Fourier components of the
periodic time-averaged density field are given by
$N\rho_{\bf G}$ where
\begin{equation}
\rho_{\bf G} = \int^{\cal V} d{\bf r} \bar{\rho}({\bf r})\exp(-i{\bf
G \cdot r}).
\label{eq42}
\end{equation}
Here ${\cal V}$ denotes the area of the unit cell. 
In thermal equilibrium, this time average is equal to a
thermal average. 

We calculate $\rho{\bf_G}\/$ by solving, 
in a one-order-parameter approximation \cite{rama},
the self-consistent equations for the
Fourier components of the periodic density field at freezing.
The input direct correlation function, $C(k_\perp,k_z)\/$, is
taken from a solution of the liquid-state equations with
a Rogers-Young closure scheme \cite{us2,roy}.
We use parameters appropriate to BSCCO i.e.
$\lambda(T=0) = 1500 - 1800 \AA$ and $d = 15 \AA$ and assume a two-fluid
temperature dependence of $\lambda$ with $T_c(0)$ = 85 K.
As mentioned earlier, the relevant input correlation function
for the freezing of the vortex liquid into an Abrikosov lattice 
is $C({\bf G}_\perp,G_z=0)\/$. In Fig.~\ref{fig6}, we plot the
Fourier components $\rho_{\bf G}\/$ of the periodic density field 
(triangles) at the freezing temperature as functions of $G$.
Note the fast decay of $\rho_{\bf G}\/$
at large $G\/$ and the nearly Gaussian envelope of the decay.
In Fig.~\ref{fig7}, we display the magnetic field distribution for
the thermally broadened Abrikosov lattice at freezing for values of
the magnetic induction of (a) $B_0$ = 0.3 T and (b) $B_0$ = 0.05T,
incorporating corrections
due to the finite core size (see Eq.(\ref{eq210})).
The field distributions obtained in this theory are notably 
narrower and more
symmetric than the corresponding distributions calculated for
an almost perfect Abrikosov lattice shown in the same figure
for comparison. This feature is seen in
experiments on BSCCO \cite{lee,lee1,harsh1}.

Although the density functional description is in principle
applicable even deep within the solid phase, it is unclear how
to parametrize the appropriate input correlation functions in
this case. While it is well-known that the existence of the
liquid-solid transition cannot be inferred from usual liquid
state methods such as the ones used by us (the correlations
show no singular behaviour), using these correlation functions
in a density functional description of the solid much below its
crystallization temperature is questionable. The issue
of how to describe the solid at temperatures far from freezing
is relevant in determining the temperature dependence of the
crystalline order parameters $\rho_{\bf G}\/$, which in turn
determines the temperature dependence of the
second moment of the field distribution function.

In this calculation, we extract the temperature dependence
of the $\rho_{\bf G}\/$'s by assuming that the 
square of the width of the equilibrium density distribution
at a lattice site is proportional to the temperature $T$ at all temperatures
lower than the melting temperature. The constant of proportionality 
is obtained by demanding that the solid at its
melting temperature has the same Lindemann parameter as
that obtained from our self-consistent density functional calculation
at freezing. A linear temperature dependence of the mean square displacement
also occurs in a harmonic treatment (with temperature-independent
elastic constants) of the fluctuations of a vortex line
about its equilibrium position at a lattice site. However, our treatment
differs from a harmonic one in that the value of the mean square displacement
at the melting temperature is obtained from our density functional
calculation which, as argued above, is non-perturbative and includes 
possible anharmonic effects.

The thermal broadening of the density distribution in the crystalline phase
is calculated in the following way. Our density functional calculations
\cite{us1,us2} show that at
fields larger than about 50 mT, the Lindemann parameter
$\sqrt{\langle u^2\rangle}/a_0\/$ 
at the freezing temperature is nearly constant at a value $\simeq$ 0.2. 
Assuming $\langle u^2(T) \rangle /a_0^2 = \chi T\/$, and using the value of
the melting temperature $(\simeq 18 K\/)$ obtained from our calculation, we
obtain $\chi\simeq 1/450 K^{-1}\/$. If we assume that the density distribution
in the crystalline state is represented by a sum of Gaussian density
profiles centered at the lattice sites, i.e.
$\bar{\rho}({\bf r})=\sum_i f({\bf r-R_i})\/$, with $f({\bf
r})=\alpha/\pi\exp(-\alpha r^2)\/$, the assumed form of $\langle u^2 \rangle$
implies that $\alpha(T) = 1/(a_0^2 \chi T)$. The time-averaged
Fourier components of the density field are then given by $\rho_{\bf G} = 
\exp(-G^2a_0^2\chi T/4)\/= \exp(-4\pi^2(m^2+n^2+mn)\chi
T/3)\/$. The second moment of the local field distribution is calculated
using 
\begin{equation}
[\Delta B^2]=B^2_0\sum_{{\bf G}\neq 0}\frac{w^2(G)K^2_1(w(G))
\rho_{\bf G} \rho_{-{\bf G}}}{(1+\lambda^2G^2)^2}.
\label{eq43}
\end{equation}
The temperature dependence of the second moment comes from two
sources -- the (assumed two-fluid) temperature
dependence of $\lambda\/$ on $T\/$, as well as the temperature
dependence of the $\rho_{\bf G}\/$s. 
Fig.~\ref{fig8} shows the temperature dependence of the 
second moment as calculated in this approximation.
Note the curvature of the plot of $(\Delta B)^2
\,vs.\, T\/$. This trend is {\it opposite} to that obtained if one
assumes a two-fluid form for the $T$-dependence of $\lambda$ and
ignores the corrections arising from the temperature-dependence of the
broadening of the density distribution. A similar trend is seen in some 
experiments \cite{harsh1} at not too high fields ($B_0 < $ 1.5 T).
While the use of Debye-Waller-like broadening factors to
rationalize the non-two-fluid
temperature dependence of $[\Delta B^2]$ is not new
(see\cite{harsh1,harsh2}), in our calculation, these factors are not
obtained as fitting parameters but follow from the general
structure of our theory.

In Fig.~\ref{fig9}, we show the temperature dependence of the
third moment, as obtained in this calculation using
\begin{equation}
[\Delta B^3]=B^3_0\sum_{\stackrel{{\bf G}_1,{\bf G}_2,{\bf G}_3 \neq
0} {{\bf G}_1+{\bf G}_2+{\bf G}_3 =0}}\frac{\sigma(G_1) \sigma(G_2)
\sigma(G_3) \rho_{{\bf G}_1} \rho_{{\bf G}_2} \rho_{{\bf G}_3}
}{(1+\lambda^2G^2_1)(1+\lambda^2G^2_2)(1+\lambda^2G^2_3)},
\label{eq44}
\end{equation}
where $\sigma(G_i)$, $i$ = 1, 2, 3 represent the cutoff correction
described in section \ref{sec2} (see Eq.(\ref{eq214})).

The results of our replica treatment \cite{disprl} of the
effects of weak pinning disorder on the correlations and 
freezing of the vortex system indicate that the results obtained 
above for the pure
system would remain essentially unchanged if the low-temperature
``solid'' state in the presence of pinning exhibits strong local
crystalline order. As discussed in section \ref{sec3}C above, 
the freezing of the liquid into such a state may be described by
a density functional (see Eq.(\ref{eq312})) in which the
quantity $[C^{(1)}({\bf r}) - C^{(2)}({\bf r})]$ plays the same
role as that of the direct correlation function $C({\bf r})$ in
the density functional theory of freezing of the pure system. Our 
replica liquid state theory \cite{disprl} shows that if the
pinning disorder is weak, then the function $C^{(1)}({\bf r})$
is nearly identical to the function $C({\bf r})$ for the pure
system, and $C^{(2)}({\bf r})$ is much smaller, so that 
$C^{(1)}({\bf r}) - C^{(2)}({\bf r}) \simeq C({\bf r})$. This,
in turn, implies that the $\rho_{\bf G}$s that describe the
(local) density distribution in the low-temperature state in the
presence of pinning are nearly the same as those obtained for
the pure system. Since the $\mu$SR spectrum in the
low-temperature state is determined by the values of the
$\rho_{\bf G}$s (see Eqs. (\ref{eq43}) and (\ref{eq44})), it  
would be nearly unaffected by pinning disorder as long as the
approximation of describing the local translational order in the
low-temperature state by the $\rho_{\bf G}$s remains valid. This
is expected because, as discussed earlier, $\mu$SR 
spectroscopy probes the {\em local} translational order of the 
vortex system.

Before we conclude this section, we comment briefly on the
liquid-solid phase boundary obtained in our calculation
in relation to the experimental data. Our calculations,
using a Rogers-Young closure scheme and a one-parameter
density functional theory, indicate that the liquid-solid phase
boundary in the pure system at high fields is practically
independent of field and occurs at $T_M \simeq 18 K$. This value is
slightly lower than the values obtained in experiments \cite{lee,lee1}.
In our calculations, we used values of $\lambda_{ab} = 1500-1800\AA$. 
Assuming larger values of $\lambda_{ab}$ would act to shift the
phase boundary to still lower temperatures. However, as 
we have argued earlier \cite{us2}, the principal source of error 
in our estimation of the phase boundary arises from the 
fact that our liquid state approximations {\it underestimate} 
correlations in the liquid. A more accurate treatment of liquid state
correlations is expected to shift the phase boundary to higher
temperatures. Also, a comparison of our results for the pure system at
high fields with the experimental data may not be particularly
meaningful because, as discussed in section \ref{sec1} above, 
the freezing observed in experiments at high fields is probably 
to a glassy state rather than to a crystalline one.

\section{$\mu$SR spectrum in a ``frozen'' liquid}
\label{sec5}

We now discuss the calculation of the moments of
$n(B)\/$ in the liquid phase of the vortex
system. If the time scale of structural relaxation in the
liquid (i.e. the time scale over which the local density becomes
homogeneous in the pure system, or nearly homogeneous if weak
pinning is present) is much smaller
than the inverse muon precession rate, the density distribution
seen by the muons should be nearly homogeneous and the line widths
vanishingly small. In the opposite limit, in which structural
relaxation in the liquid occurs over times that are much longer 
than the inverse muon precession rate and the muon lifetime, some
quantitative progress may be made in calculating $\mu$SR lineshapes and
linewidths. We propose that such a limit is appropriate for a
description of the low-temperature, high-field state of the 
flux line ensemble in BSCCO. 
Such glassy behavior may have several origins, as discussed
in the Introduction. Irrespective of the precise nature
of this state, it is a reasonable guess that it
exhibits short range positional correlations but no long range
crystalline order. A natural assumption then, is that positional
correlations in this state should resemble those in 
a liquid. As mentioned in the Introduction, this proposal
is consistent with existing neutron scattering \cite{neutr} and
other experimental data. It is also physically sensible, as it 
provides a way by which some degree of local correlations can be
built into the glassy state, while ensuring that crystalline long-range
order is destroyed, as required by the Larkin-Ovchinnikov 
argument \cite{larkin}. In this section, we explore the 
consequences of this proposal.

In this (adiabatic) limit, the muons essentially see a
``snapshot'' of the liquid (i.e. a typical liquid-like configuration of
the pancake vortices). The second moment of the field distribution
function is then given by the {\em space-averaged} two-point
correlation function of the density in the configuration seen by
the muons. If the system is self-averaging (we assume that it is
so), then this correlation function is nothing but the {\em
time-averaged} (equilibrium) two-point density correlation
function of the liquid. Therefore, the linewidth in this limit
can be calculated in terms of the static structure factor
$S(k)\/$ of the liquid, which is defined as
\begin{equation}
S({\bf k}) = \frac{1}{N}\langle\rho_{\bf k}\rho_{-{\bf k}}\rangle = 
1+\frac{1}{N}\langle\sum_{i \neq j} \exp(i{\bf k \cdot r}_{ij})\rangle,
\label{eq51}
\end{equation}
where ${\bf r}_{ij} = {\bf r}_i - {\bf r}_j$ (${\bf r}_i$ is the
position vector of the $i$th pancake vortex), $N$ is the total number
of vortices, the sum runs over all the vortices, and the
brackets $\langle\cdots\rangle$ denote a thermal average.
The second moment of the field distribution function is given by
\begin{equation}
[\Delta B^2]=[\Delta B_\perp^2]+[\Delta B_z^2].
\label{eq52}
\end{equation}
The brackets $[\cdots]$ denote a space average for a typical
liquid-like configuration of the vortex system. Precisely
which moment ($[\Delta B_z^2]$ or $[\Delta B_\perp^2]$) is
measured depends on the geometry in which the
experiments are done. Our results are illustrated here for
the transverse $\mu$SR geometry, in which the muons
are injected with their polarizations in the plane of the
superconducting layers and thus perpendicular to the
external field. In this geometry, the measured quantity
is $[\Delta B_z^2]$.

The equivalence of space and thermal averages then leads to the
following expression for the second moment of the distribution of 
$B_z$:
\begin{equation}
[\Delta B_z^2]=\frac{\rho_\ell}{d}\frac{1}{(2\pi)^3}\int
d^2k_\perp dk_z b^2_z(k_\perp,k_z) S(k_\perp,k_z),
\label{eq53}
\end{equation}
where $\rho_\ell \equiv B_0/\Phi_0$ denotes the (areal) density
of vortices, and the quantity $b_z$ is defined in Eq.({\ref{eq33}).
A similar expression for $[\Delta B^2_\perp]$ may be obtained by
replacing $b_z$ by $b_\perp$ (see Eq.(\ref{eq32})) in the above
equation. 

Before analyzing the relevance of short-range correlations in
the computation of moments of $n(B)\/$ in the frozen liquid, we
discuss two simple limiting cases. First consider the limit 
in which the liquid has {\it no} correlations, so that 
$S(k_\perp,k_z)=1\/$, and the second
moment of the distribution of the $z$-component of the field is
given by 
\begin{equation}
[\Delta B_z^2]=\frac{\rho_\ell}{d}\frac{\Phi_0^2d^2}{(2\pi)^2}\int_0^\infty
k_\perp dk_\perp \int_{-\infty}^{\infty}
\frac{dk_z}{(1 + \lambda^2k_\perp^2 +
\lambda^2k_z^2)^2} = B_0\Phi_0d/8\pi\lambda^3. 
\label{eq54}
\end{equation}
For $B=0.5$T, $d = 15\AA$ and $\lambda = 1500\AA$, we obtain
$[\Delta B_z^2]$ = 1830 G$^2$, whereas assuming $\lambda = 1800
\AA$ gives $[\Delta B_z^2]$ = 1059 G$^2$.

Now consider the case of a completely
uncorrelated arrangement of rigid flux lines.
The second moment of the field distribution is then given by
\begin{equation}
[\Delta
B^2_z]=\frac{\rho_\ell}{(2\pi)^2}\int\frac{\Phi^2_0}{(1+\lambda^2
k^2_\perp)^2}d^2k_\perp =\frac{B_0\Phi_0}{4\pi\lambda^2}.
\label{eq56}
\end{equation}
Assuming, as before, $B_0$ = 0.5T and $\lambda = 1500\AA$, we
obtain $[\Delta B_z^2]$ = 3.66$\times 10^5$ G$^2$ (2.54 $\times
10^5$ G$^2$ for $\lambda = 1800 \AA$). These numbers are to be
compared with the value of $[\Delta B_z^2]$ (= 3140 G$^2$ for
$\lambda = 1500\AA$, 1514 G$^2$ for $\lambda = 1800\AA$) for a
perfect Abrikosov lattice.

These simple results, first derived
by Brandt \cite{brandt1}, bring out an important point: the
destruction of correlations in the $ab$-plane (while keeping
the vortices perfectly correlated in the $c$-direction) greatly
{\em increases} the linewidth, whereas the destruction of
correlations in the $c$-direction causes the linewidth to become
smaller. This observation suggests that a likely candidate for a
structure that exhibits anomalously narrow $\mu$SR
linewidths would be one with at least local (see below) positional
correlations in the layer planes and little correlation across
the layers. Our proposal for the structure of the ``glassy''
state of the vortex system is consistent with this expectation:
our liquid state calculations \cite{us1,us2} show that the
equilibrium liquid just above freezing exhibits pronounced
short-range correlations in the layer plane, but the correlation
length in the $c$-direction remains small (about 15 -- 20 layer
spacings). 

We now describe a simple calculation that illustrates that the
presence of {\em short-range} $ab$-plane correlations in a frozen
liquid of rigid flux lines is sufficient to
bring down the the linewidth to values comparable to that
obtained for a perfect Abrikosov lattice.
The inclusion of local correlations in the calculation of the
moments of the field distribution requires the specification of
the liquid structure factor $S(k_\perp)\/$. At the simplest
level, $S(k_\perp)\/$ can be approximated as
\begin{eqnarray}
S(k_\perp)&=& 0 \;~~~ (0 < k_\perp < k_m-\epsilon), \nonumber \\
&=& S_m \; ~~~ (k_m-\epsilon <  k_\perp < k_m+\epsilon), \nonumber \\
&=& 1 \; ~~~ (k_m+\epsilon < k_\perp < \infty).
\label{eq57}
\end{eqnarray}
This approximation retains all the basic qualitative features
expected of a fluid structure factor -- it rises
from a value of zero near the origin of $k-$space to a peak at
a $k-$space distance of the order of $2\pi$ times the inverse of the mean
interparticle spacing, and becomes unity at sufficiently large $k\/$.
With this approximation, it is easy to obtain 
\begin{equation}
[\Delta B^2_z]=\frac{B\Phi_0}{2\pi\lambda^4}\left(\frac{2\epsilon
S_m}{k_m^3} +\frac{1}{2k_m^2}\right). 
\label{eq58}
\end{equation}
Assuming $2 \epsilon S_m/k_m \sim 0.5\/$, so that $[\Delta B^2_z]\simeq
B\Phi_0/(2\pi\lambda^4k_m^2)\/$, $k_m^2\simeq 4 \pi^2 B/\Phi_0\/$, and
for sufficiently large fields (so that $(1+\lambda^2k^2_m)
\simeq \lambda^2k^2_m\/$), we obtain
\begin{equation}
[\Delta B^2_z]\simeq \frac{\Phi^2_0}{8 \pi^3 \lambda^4} \simeq 4
\times 10^{-3} \frac{\Phi_0^2}{\lambda^4},
\label{eq59}
\end{equation}
which is quite close to the result, $[\Delta B^2_z]\simeq 3.71
\times 10^{-3} \frac{\Phi_0^2}{\lambda^4}$, for the linewidth in
a perfect Abrikosov lattice.  This admittedly crude calculation
thus brings out the important point that the $\mu$SR linewidth is
sensitive only to the {\em local} order.

We now proceed to describe our detailed calculation of the
linewidths in a frozen liquid-like state. In terms of the direct
correlation function, Eq.(\ref{eq53}) can be written as 
\begin{equation}
[\Delta B_z^2]=\frac{\rho_\ell}{d}\frac{1}{(2\pi)^2}\int^{\infty}_0
k_\perp dk_\perp
\int^{\infty}_{-\infty}dk_z 
\frac{\phi^2_0d^2}{(1+\lambda^2k^2_\perp+\lambda^2k^2_z)^2}
\frac{1}{(1-\rho_\ell C(k_\perp,k_z))},
\label{eq510}
\end{equation}
where $C(k_\perp,k_z)\/$ is the liquid direct correlation
function and we have used the relation 
$S(k_\perp,k_z)=1/(1-\rho_\ell C(k_\perp,k_z))\/$. 
The direct correlation function $C(k_\perp,k_z)\/$ is obtained
from our earlier calculation \cite{us1,us2} of the
equal-time correlations in a layered vortex liquid. To represent
the core form factors (in order to reduce the complexity of the
expressions, we do not explicitly show these form factors in the
equations derived below), we model the core of a pancake vortex 
as a cylinder of radial dimension $\xi_{ab} \simeq 15 \AA$
and height $\xi_c \simeq 3\AA$. We use the decomposition
\begin{eqnarray}
C(\rho,nd)=C_s(\rho,nd)+C_L(\rho,nd),
\label{eq511}
\end{eqnarray}
where the ``long-range'' part of $C$ is given by 
$C_L(\rho,nd)=-\beta V(\rho,nd)$ ($V(\rho,nd)$, the
intervortex interaction in position space, is the Fourier transform
of $V({\bf k})$ defined in Eq.(\ref{eq31})).  Asymptotically, $C(\rho,nd)$ tends to the value
$-\beta V(\rho,nd)$, which we have called $C_L(\rho,nd)$, 
so that $C_s(\rho,nd)$ tends to zero. This
``short-range'' part is assumed to have the form
$C_s(\rho,nd)=C_s(\rho,n=0) \delta_{n,0}\/$ (see
\cite{us1,us2}). The Fourier transform, $C_s(k_\perp)$, of 
the function $C_s(\rho,n=0)$ is obtained by numerically
solving the self-consistent equations of liquid state theory.
Approximating $\sin^2(k_zd/2)\/$ appearing in Eq.(\ref{eq31}) 
by $k^2_zd^2/4\/$, the integral over $k_z\/$ in Eq.(\ref{eq510})
can be performed to yield
\begin{equation}
[\Delta B^2_z]=\frac{\Phi_0^2 d}{4\pi \Gamma\lambda^2}
\int^\infty_0
dk_\perp k_\perp^3\left(\sqrt\frac{Y}{X}
- \sqrt{\frac{\lambda^2}{1 + \lambda^2 k_\perp^2}}\right),
\label{eq512}
\end{equation}
where
\begin{eqnarray}
X&=&k^2_\perp(1+\lambda^2k^2_\perp)(1-\rho_\ell C_s(k_\perp))
+\rho_\ell\Gamma\lambda^2k^2_\perp, \nonumber \\
Y&=&\lambda^2k^2_\perp(1-\rho_\ell
C_s(k_\perp))+\rho_\ell\Gamma\lambda^2.
\label{eq513}
\end{eqnarray}
These equations and the numerically obtained $C_s(k_\perp)$ are
used to evaluate the second moment $[\Delta B^2_z]$.

The third moment of the field
distribution function in the liquid phase can be obtained
via the liquid state methods we have discussed here, together
with a simple decoupling approximation. Using
\begin{equation}
[\Delta B_z^3] = \frac{1}{V^3} \sum_{{\bf k}_1,{\bf k}_2,{\bf k}_3}
\langle\rho({\bf k}_1)\rho({\bf k}_2)\rho(-{\bf k}_1-{\bf k}_2)\rangle
b_z({\bf k}_1)b_z({\bf k}_2)b_z(-{\bf k}_1-{\bf k}_2),
\label{eq518}
\end{equation}
what is required to calculate this moment 
is the triplet structure factor $S^{(3)}({\bf k}_1,{\bf k}_2)\/$ defined by
\begin{equation}
S^{(3)}({\bf k}_1,{\bf k}_2) = \frac{1}{N}\langle\rho({\bf k}_1)\rho({\bf k}_2)
\rho(-{\bf k}_1-{\bf k}_2)\rangle.
\label{eq519}
\end{equation}
In terms of the triplet direct
correlation function $C^3({\bf k}_1,{\bf k}_2)$,
this assumes the form
\begin{equation}
C^3({\bf k}_1,{\bf k}_2)=1-\frac{S^{(3)}({\bf
k}_1,{\bf k}_2)}{S({\bf k}_1)S({\bf
k}_2)S(-{\bf k}_1-{\bf k}_2)}.
\label{eq520}
\end{equation}
The triplet correlation function is typically small in magnitude
in regular three dimensional solids. If it is assumed to be
zero, (the so-called convolution relation \cite{convolution}), the
integral can be written completely in terms of the
the static structure factor. We thus obtain
\begin{eqnarray}
[\Delta B_z^3] = \frac{\rho_\ell}{d}\frac{1}{(2\pi)^6}\int d{\bf k}_{1\perp}
\int dk_{1z}\int d{\bf k}_{2\perp}\int dk_{2z} \nonumber \\
S({\bf k}_1)S({\bf k}_2)S(-{\bf k}_1-{\bf k}_2)
b_z({\bf k}_1)b_z({\bf k}_2)b_z(-{\bf k}_1-{\bf k}_2).
\label{eq521}
\end{eqnarray}

This expression is defined in terms of the
two-particle structure factor $S(k_\perp,k_z)$ which, as
discussed above, may be obtained from the function
$C_s(k_\perp)$ calculated in our liquid state theory.
Exploiting the structure of this expression, and after
considerable algebra, the 6-dimensional
integral is reduced to a 3-dimensional one which is evaluated
numerically. 

Our results for $[\Delta B_z^2]$ and $[\Delta B_z^3]$
for values of $\lambda_{ab} = 1500 \AA$
and $1800\AA$ are summarized in Table I. These results are
obtained using the $C_s(k_\perp)$ calculated just above 
the melting transition.
The core form factors, omitted for simplicity in
the expressions derived above, are present in the
full calculation. While our liquid-state calculations are performed for
$\lambda = 1500\AA$, the calculated $C_s(k_\perp)$ 
can be used for the case $\lambda_{ab} = 1800\AA$
in the high field limit provided the melting temperature is scaled
as $T_M \propto 1/\lambda_{ab}^2$. This is a 
simple consequence of the crossover to quasi-two-dimensional
behavior at high fields, in which the dominant
interaction is the (scale-invariant) logarithmic
coupling between vortices on the same plane, which
is perturbed only weakly by the presence of vortices
on other planes. Comparison with experimental data \cite{data} 
shows that our proposal agrees quite well with the results
of experiments.

Results for $[\Delta B_\perp^2]$ can be obtained from
\begin{equation}
[\Delta B^2_\perp]=\frac{\Phi_0^2 d}{4\pi \Gamma\lambda^2}
\int^\infty_0
dk_\perp k_\perp\left[
\sqrt{\frac{1 + \lambda^2 k_\perp^2}{\lambda^2}}
- \sqrt{\frac{X}{Y}}\right].
\label{eq517}
\end{equation}
Here $X$ and $Y$ are as defined in Eq.(\ref{eq513}). The value of this
moment is comparable to $[\Delta B_z^2]$.

Our {\it ansatz} that the vortices remain stationary during the
lifetime (or precession time if it is smaller) of the muons
is expected to be strictly valid only at very low temperatures -- at
higher temperatures we would expect the effects of 
thermal broadening of the (now inhomogeneous) 
density distributions to {\it reduce} linewidths further
in analogy with our results for the crystalline
phase in the pure system. There seems to be no simple
way of incorporating such thermal effects into the
theory described here (see, however, the discussion in section
\ref{sec7} below). A theoretical investigation of the
temperature dependence of the linewidth in this disordered phase
would be interesting. It would also be worthwhile to carry out a
more accurate calculation of the third moment of the
field distribution by going beyond the simple decoupling
approximation used here.

In the calculations described above, we used the equilibrium
correlation function of the {\it pure} liquid at a temperature
just above the crystallization temperature $T_M$ to determine 
the moments of the field
distribution. Since the glassy phase we are interested in is
expected to arise as a consequence of the presence of pinning
disorder, one should, in principle, use the correlation function
of the disordered liquid near its freezing (glass transition) 
temperature in these calculations. Due to reasons mentioned
below, we expect that the results of such a calculation would be
essentially the same as those described above. Our replica
calculation \cite{disprl} of liquid state correlations in the
presence of pinning disorder shows that the equal-time
(equilibrium) density correlations in the liquid are basically
unaffected by weak disorder. Therefore, our approximation of
replacing the static structure factor of the disordered liquid
by that of the pure one is expected to be quite accurate.
We also find that the value of the linewidth depends relatively 
weakly on the temperature at which the liquid-state correlations
are calculated -- a change in the  temperature from $T_M$ to
$3T_M$ leads to a decrease of about $20\%$ in the
linewidth $[\Delta B^2]$. This observation indicates that a
small difference between $T_M$ and the actual glass transition
temperature of the disordered liquid would not cause any
significant change in the final results.

\section{$\mu$SR linewidth in the disordered liquid}
\label{sec6}

The time scales for structural relaxation in the high-temperature 
liquid phase are expected to be short.
The second moment of the field distribution function in this  
phase of the pure system would be vanishingly small if these
time scales are fast compared to the inverse precession rate of the muons. 
Even in this limit, however, measurable line-widths can obtain in the
presence of quenched pinning disorder. The source of this
broadening is the off-diagonal (in replica space) density correlation
function. Off-diagonal correlations, which are absent in the
pure system, become non-zero in the presence of disorder. These
correlations arise due to the disorder-induced inhomogeneity of
the time-averaged local density. In a replica symmetric description, 
the off-diagonal radial distribution function is defined as 
\begin{equation}
g^{(2)}({\bf r}) \equiv 
\langle\langle\rho^\alpha(0)\rho^{\beta\neq\alpha}({\bf r})\rangle\rangle/\rho_\ell^2
= [\langle\rho(0)\rangle\langle\rho({\bf r})\rangle]/\rho_\ell^2\/.
\label{eq61}
\end{equation}
Here the $\rho$s represent
local number densities, $\alpha$ and $\beta$ are replica indices,
$\langle\langle\cdots\rangle\rangle$ represents an average over
the canonical distribution function of the replicated system and
$\langle\cdots\rangle\/$, a thermal average for the original disordered
system prior to averaging over disorder. The square brackets
$[\cdots]\/$ denote an average over the probability
distribution governing the disorder. 

If the relaxation times of
the liquid are short, then the muons see the time-averaged
(equilibrium) desnsity distribution which is inhomogeneous if
pinning disorder is present. In this case, the 
second moment of the field distribution function can be obtained
from the correlation function $h^{(2)}\equiv g^{(2)}-1$ that describes the
correlation of the deviations of the time-averaged local density
from its space-averaged value $\rho_\ell$. Specifically, 
the second moment of the distribution
of the $z$-component of the local magnetic field is given by
\begin{equation}
[\Delta B^2_z]=\frac{\rho_\ell}{d}\frac{1}{(2\pi)^3}\int
d{\bf k}_\perp \int dk_z b^2_z S^{(2)}(k_\perp,k_z),
\label{eq62}
\end{equation}
where $S^{(2)}({\bf k})\/$, the off-diagonal structure factor
defined as $[\langle\delta\rho_{\bf k}\rangle\langle\delta\rho_{-{\bf k}}\rangle]/N\/$,
is given by (see Eq.(\ref{eq311}))
\begin{equation}
S^{(2)}({\bf k}) = \rho_\ell h^{(2)}({\bf k}) = 
\frac{\rho_\ell C^{(2)}({\bf k})}{[1-\rho_\ell C^{(1)}({\bf k})
+ \rho_\ell C^{(2)}({\bf k})]^2}.
\label{eq63}
\end{equation}
Using this and the expression, Eq.(\ref{eq33}), for $b_z$,
Eq.(\ref{eq62}) can be written as
\begin{eqnarray}
[\Delta B^2_z]&=&\frac{\rho_\ell}{d}\frac{1}{(2\pi)^3}\int
d{\bf k}_\perp \int d k_z
\frac{\Phi^2_0d^2}{(1+\lambda^2k^2_\perp+\lambda^2k^2_z)^2}
\frac{\rho_\ell C^{(2)}(k_\perp,k_z)}{[1-\rho_\ell C^{(1)}(k_\perp,k_z)
+ \rho_\ell C^{(2)}(k_\perp,k_z)]^2}.
\label{eq64}
\end{eqnarray}
Separating the diagonal direct correlation function $C^{(1)}$ as
before into a short-range and a long-range part and
evaluating the integral over $k_z\/$, we get
\begin{equation}
[\Delta B^2_z]=\frac{\rho_\ell}{d}\frac{\Phi_0^2 d^2}{8\pi}
\int^\infty_0 k_\perp dk_\perp \rho_\ell
C^{(2)}(k_\perp)\frac{1}{\sqrt{P}Q^{3/2}},
\label{eq65}
\end{equation}
where $P$ and $Q$ are given by
\begin{eqnarray}
P&=&(1+\lambda^2k^2_\perp)(1-\rho_\ell C^{(1)}_s(k_\perp)
+ \rho_\ell C^{(2)}(k_\perp)) + \rho_\ell\Gamma\lambda^2,\nonumber \\
Q&=&\lambda^2(1-\rho_\ell C^{(1)}_s(k_\perp) + \rho_\ell C^{(2)}(k_\perp))
+\frac{\rho_\ell\Gamma\lambda^2}{k_\perp^2}.
\label{eq66}
\end{eqnarray}
In deriving the above equation, we have used the approximation
$C^{(2)}(\rho,nd) = C^{(2)}(\rho)\delta_{n,0}$, so that its
Fourier transform $C^{(2)}(k_\perp,k_z)$ does not depend on
$k_z$. This approximation is justified \cite{disprl} because the interaction
between different replicas (Eq.(\ref{eq318})) does not couple
vortices on different layers.

Evaluating the integral in Eq.(\ref{eq65}) numerically, using the values for
$C^{(1)}$ and $C^{(2)}$ obtained in Ref.\cite{disprl} through
a self-consistent solution of the replicated liquid state
equations, we obtain values for $[\Delta
B^2_z]\/$ in the range $1-10 G^2\/$. At $B_0$ = 0.3 T, just above the
melting transition, our calculations yield $[\Delta B^2_z] = 4.7 G^2\/$.
The smallness of this number reflects the numerically
small value of $C^{(2)}(k_\perp)$. Further, this quantity vanishes as 
the off-diagonal coupling (equivalently, the disorder strength) 
is reduced to zero, leading to the expected vanishing of the
linewidth in the high-temperature liquid phase of the pure
system. The relatively strong dependence of
$C^{(2)}$ on the disorder strength suggests that the linewidths obtained 
in the liquid phase at high temperatures should increase as 
the sample is progressively made more disordered. 

\section{Summary and discussion}
\label{sec7}
In this paper we have presented a theory for
$\mu$SR spectra in anisotropic, layered
superconductors which incorporates thermal effects in both
the solid and the liquid phase, and the effects of weak pinning
disorder. In particular, we have computed the effects of thermal fluctuations
and disorder in determining the $\mu$SR linewidths and
lineshapes in the mixed phase of the extensively studied 
high-temperature superconductor BSCCO. 
In the crystalline solid phase, our
density functional description of the time-averaged density
distribution offers a {\it nonperturbative} way of
assessing the effects of thermal broadening. Our theory
of $\mu$SR spectra in the liquid phase represents, to the
best of our knowledge, the first attempt to calculate linewidths
in this phase beyond the simplest assumption of
uncorrelated lines or vortices. Our analysis suggests
that the experimentally obtained linewidths in
the low-temperature, high-field state in BSCCO may be understood
if one assumes that vortices in this regime are
in a frozen state that resembles a liquid in its
local correlations. We have also quantified the effects of
pinning disorder on the $\mu$SR linewidth in the
high-temperature vortex liquid. 
The results of our calculations are in
qualitative (quantititative in some cases) agreement with those
obtained in experiments on BSCCO.

An {\it implicit} assumption we have made is
that the flux-lattice can be treated classically, i.e. the
characteristic temperature below which quantum effects are
non-negligible lies far below the equilibrium melting
temperature. This assumption has been questioned recently
by Bulaevskii {it et al.} \cite{bul} who argue
that quantum effects are non-negligible except at temperatures close to
the superconducting transition temperature $T_c(0)\/$. Their
argument rests on the assumption that the characteristic
energy scale for quantum effects is set by the superconducting 
gap and therefore is $\sim k_B T_c(0)\/$. While much is still to
be understood about the
precise role of quantum effects in the mixed phase, we
believe that the success of the density functional approach
in describing flux-lattice melting (which assumes that a
classical liquid state description can be used in treating
the liquid) serves as a counter example to the
assertion that quantum effects must be included in any
theory for the mixed phase. As we have emphasized, our
theory yields a value of the freezing temperature that is lower
than the experimentally observed value -- quantum effects
would surely act to depress this value further. 

Our results for the effects of pinning disorder on the $\mu$SR
spectrum are based on the replica liquid state theory developed
in Ref.\cite{disprl} where only replica symmetric solutions of
the liquid state equations were considered. A recent study
\cite{parisi} of the same set of equations for simple isotropic liquids
shows that these equations exhibit a replica symmetry breaking
transition which may be interpreted as a glass transition. This
work suggests that the low-temperature, high-field 
state of BSCCO may correspond to a replica symmetry
broken solution of the liquid state equations derived in
Ref.\cite{disprl}. In Ref.\cite{parisi}, the two-point correlation 
function of the ``frozen'' local density in the glassy phase is 
found to be quite similar to the equal-time (equilibrium) correlation 
function of the local density in the liquid phase. This observation
supports the {\it ansatz} we have used in this paper for the
structure of the low-temperature, high-field state of BSCCO.\\

\noindent{\bf Acknowledgements:}\\
One of us (GIM) is grateful for support from NSERC of Canada.

\begin{table}
\label{table1}
\caption{Summary of our results for $[\Delta B_z^2]$ and
$[\Delta B_z^3]$ in a slowly relaxing liquid of pancake vortices
just above its equilibrium freezing temperature $T_M$. The
results shown are for $B_0$ = 0.3 T, and two values (1500 $\AA$
and 1800 $\AA$) of the $ab$-plane penetration depth $\lambda$.
According to our {\it ansatz} (see text) for the structure of
the low-temperature disordered phase found in the high-field 
mixed-phase regime of BSCCO, these calculated values should 
correspond to the experimentally observed data in this phase at
low temperatures. Experimental $\mu$SR spectra for BSCCO crystals 
at 5 K show, for $B_0$ = 0.3, 0.4 and 1.5T respectively,
squared linewidth values of $[\Delta B^2]$ = 90 $\pm$ 4, 80 $\pm$ 5,
and 52 $\pm$ 8 G$^2$ and third moments of $[\Delta B^3]$ =
1000 $\pm$ 80, 500 $\pm$ 100 and 0 $\pm$ 250
G$^3$ (see Ref.\protect\cite{harsh1,harsh2}.}

\begin{tabular}{|c|c|c|c|}
\multicolumn{1}{|c|}{~~$\lambda (\AA)$~~}&\multicolumn{1}{c|}
{~~$[\Delta B_z^2]$ (G$^2)$~~}&
\multicolumn{1}{c|}{~~$[\Delta B_z^3 ]$ (G$^3)$~~}&\multicolumn{1}{c|}
{~~$T_M$ (calculated)~~}\\ \hline
1500 & 175.3 & 4022.0 & 
$\simeq$ 18 K \\ \hline
1800 & 111.9 & 1117.5 & 
$\simeq$ 12 K \\
\end{tabular}

\end{table}
\begin{figure}[htb]
\label{fig1}
\centerline{
\hbox {
        \epsfxsize=11.8cm
        \epsfbox{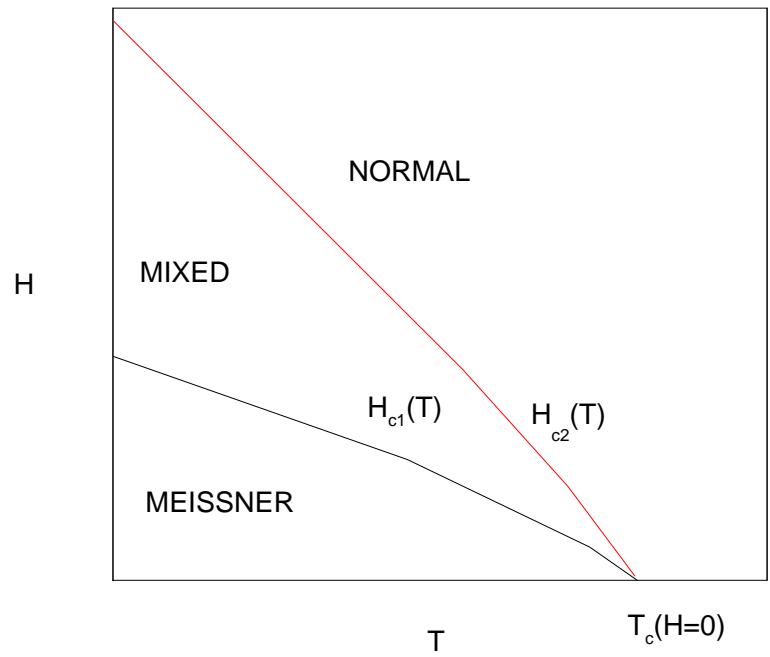}
        \vspace*{0.5cm}
       }
       }     
\caption{Schematic $H-T$ phase diagram of a type-II superconductor
in mean-field theory, illustrating Meissner, mixed and
normal phases separated by continuous phase transitions.}
\end{figure}

\begin{figure}[htb]

\centerline{
\hbox {
        \epsfxsize=11.8cm
        \epsfbox{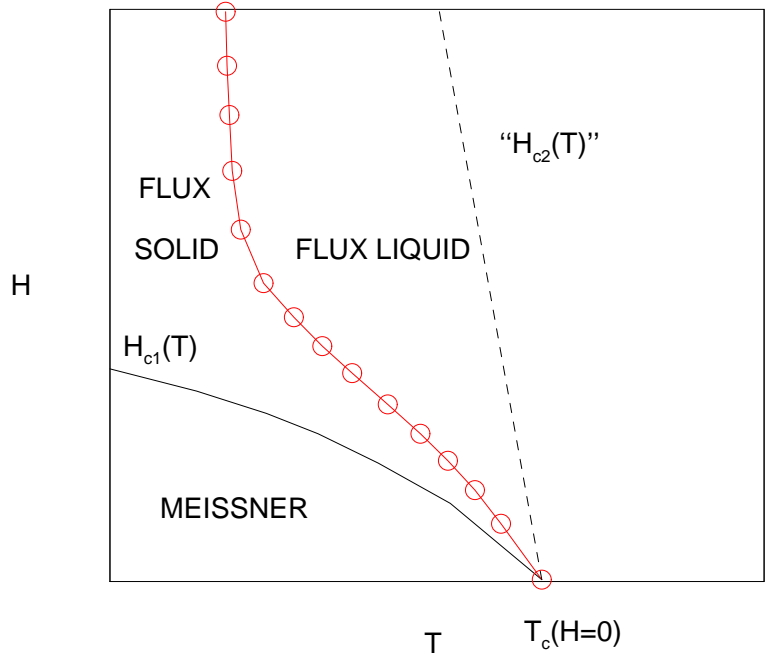}
        \vspace*{0.5cm}
       }
       }     
\caption{Schematic $H-T$ phase diagram for pure 
BSCCO illustrating the effects of thermal fluctuations on 
the mixed phase. Note the large region of ``flux liquid'' phase 
separated from a low-temperature
crystalline phase by a first-order liquid-solid
phase boundary. A very narrow region of liquid phase expected
just above the the $H_{c1}$ line is not shown. The mean-field 
$H_{c2}$ curve represents a crossover between a regime with strong
amplitude fluctuations and one in which amplitude fluctuations
are very weak.}
\label{fig2}
\end{figure}
\begin{figure}
\centerline{
\hbox {
        \epsfxsize=11.8cm
        \epsfbox{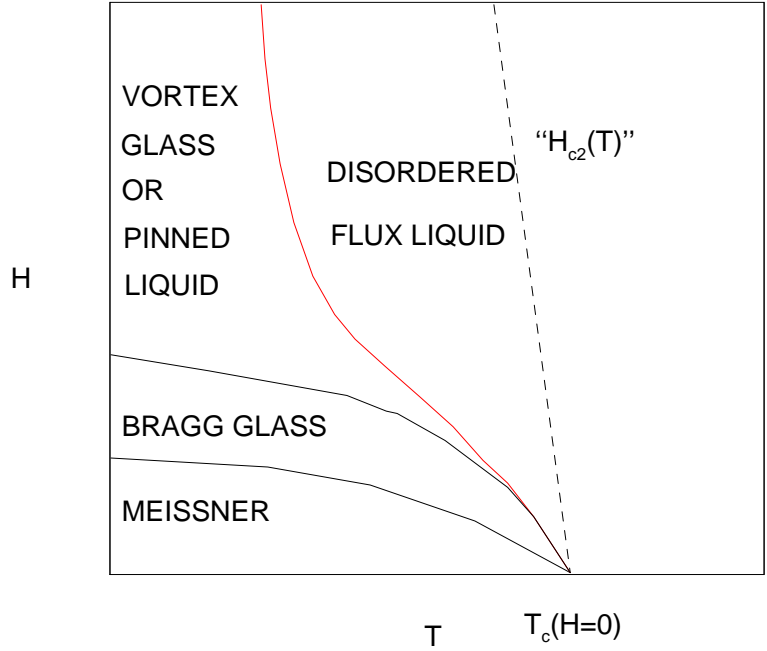}
        \vspace*{0.5cm}
       }
       }     
\caption{Schematic phase diagram for BSCCO
incorporating effects of thermal fluctuations and quenched
disorder due to random point pinning. 
The phases shown are the ``Bragg Glass''
phase (I), the ``pinned liquid'' or ``vortex glass'' phase (II),
the disordered liquid phase (III) and the normal phase (IV). The existence
of the Bragg glass and vortex glass phases has not been
established conclusively. The boundaries shown between phases
(I) and (II) and between phases (II) and (III) may actually correspond
to rapid crossovers instead of true phase transitions.}
\label{fig3}
\end{figure}

\begin{figure}
\centerline{
\hbox {
        \epsfxsize=11.8cm
        \epsfbox{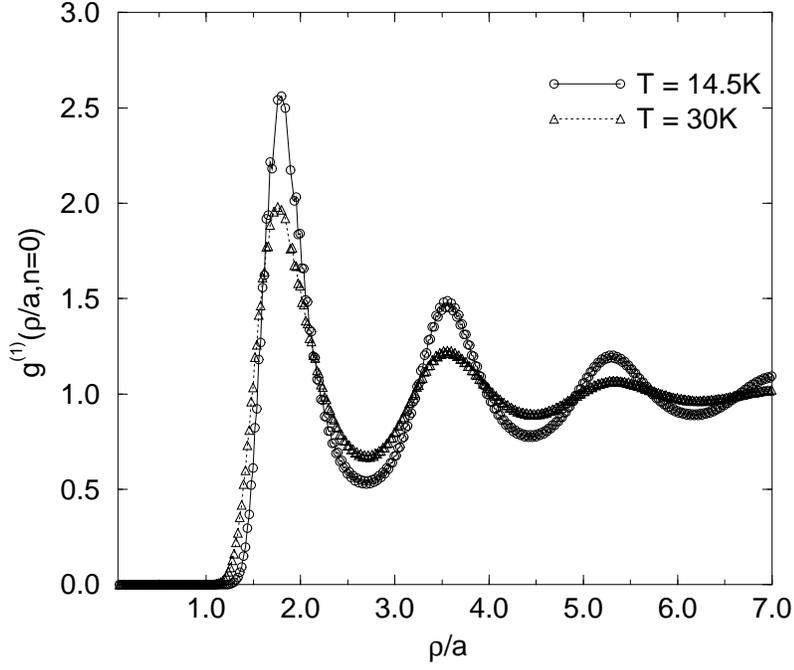}
        \vspace*{0.5cm}
       }
       }     
\caption{The diagonal correlation function 
$g^{(1)}(\rho,n=0)$, obtained using the formalism
described in section IIIC, is plotted as a function
of the in-plane separation $\rho/a$, where $a$ is defined by
$\pi a^2 = \Phi_0/B_0$. The results shown are for 
disorder strength $\Gamma^\prime = 3 \times 10^{-5} \Gamma^2$
(see text for the definitions of $\Gamma$ and $\Gamma^\prime$),
$B_0$ = 0.3 T, and $T$ = 14.5 K and 30 K.}
\label{fig4}
\end{figure}

\begin{figure}
\centerline{
\hbox {
        \epsfxsize=11.8cm
        \epsfbox{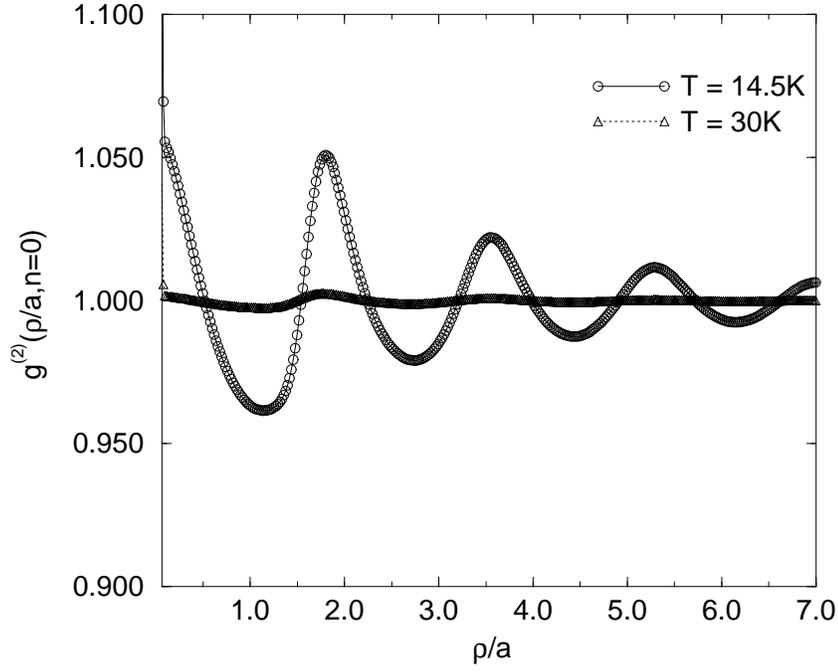}
        \vspace*{0.5cm}
       }
       }     
\caption{The off-diagonal correlation function
$g^{(2)}(\rho,n=0)$, obtained using the formalism
described in section IIIC, is plotted as a function
of the in-plane separation $\rho/a$, where $a$ is defined by
$\pi a^2 = \Phi_0/B_0$. 
The parameter values are the same as those in Fig.~4.}
\label{fig5}
\end{figure}

\begin{figure}
\centerline{
\hbox {
        \epsfxsize=11.8cm
        \epsfbox{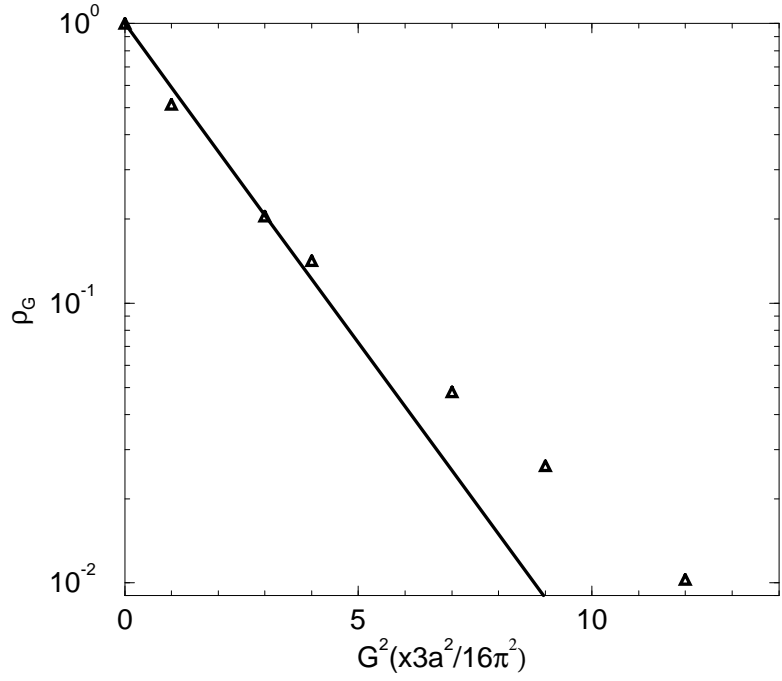}
        \vspace*{0.5cm}
       }
       }     
\caption{Fourier components $\rho_{\bf G}\/$ of the periodic density field 
at the freezing temperature, as
calculated in the density functional theory
(triangles), plotted versus $|{\bf G}|^2\/$ on a
semi-logarithmic scale. The solid line corresponds to a Gaussian
form for $\rho_G\/$ (see text) which has the same Debye-Waller factor
as that obtained in the density functional theory at the freezing
transition.} 
\label{fig6}
\end{figure}

\begin{figure}[htb]
\centerline{
\hbox {
      \vspace*{0.5cm}
        \epsfxsize=9.3cm
        \epsfbox{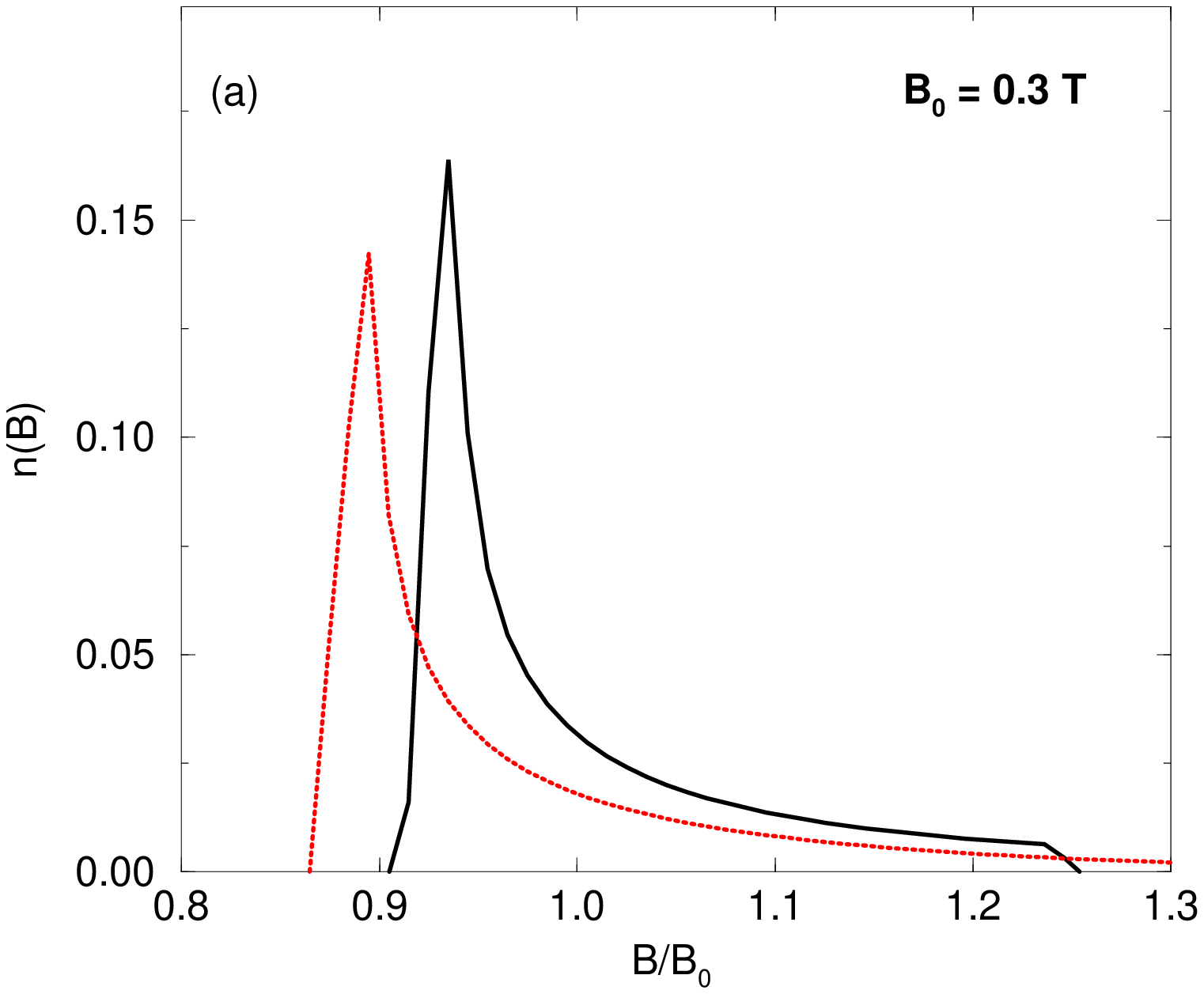}
        \epsfxsize=9.3cm
        \epsfbox{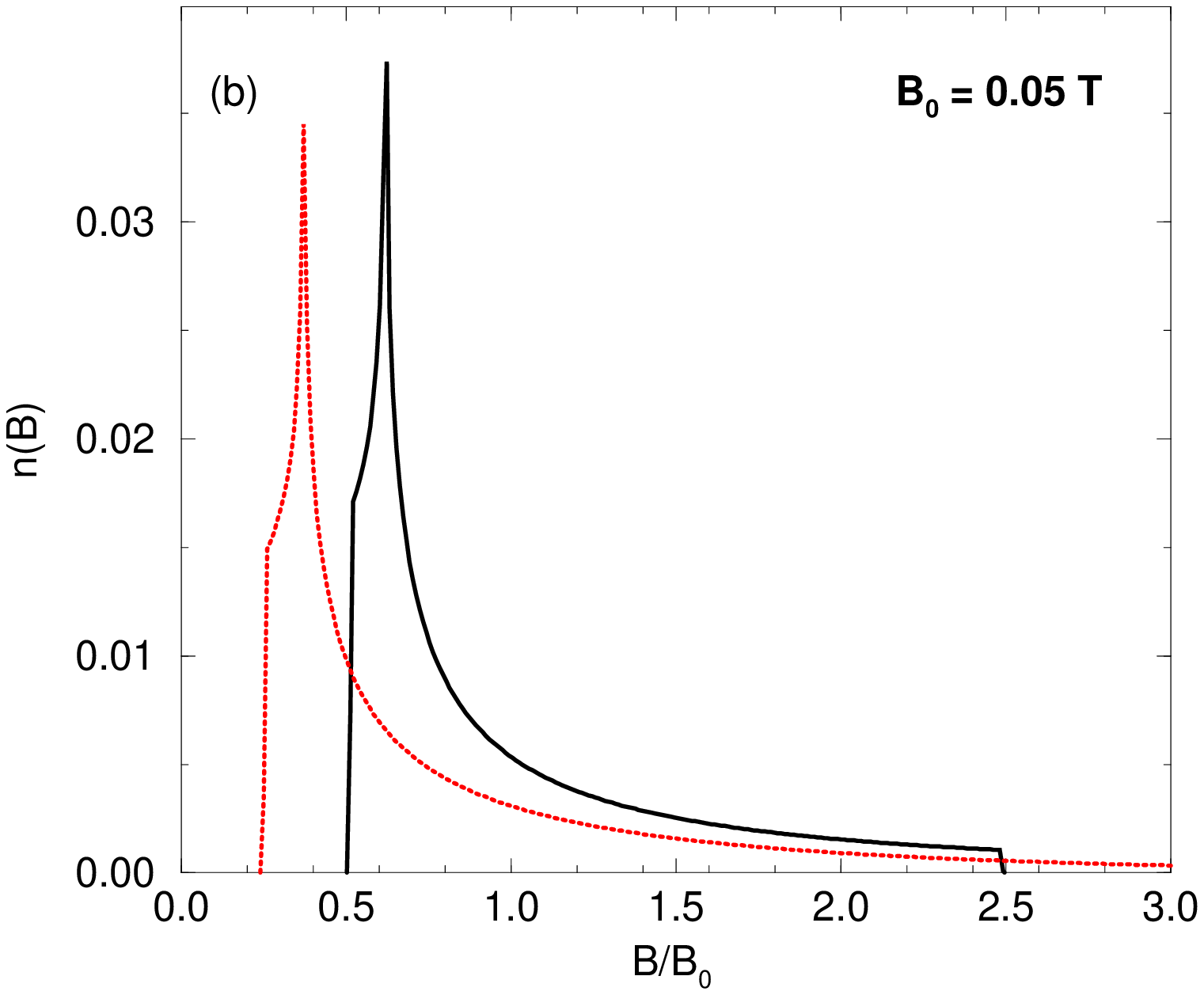}
       }
          }     
\caption{Magnetic field distribution function $n(B)\/$ in the
crystalline phase at the freezing temperature for (a) $B_0$ = 0.3 T
and (b) $B_0$ = 0.05T.
The thermally broadened time-averaged density distribution in the
crystalline state is obtained using the results of
the density functional theory. The distribution $n(B)$ 
for a slightly broadened Abrikosov flux-lattice (dashed lines) is
also shown for comparison. Note that the distribution given by
the density functional calculation is considerably narrower
and more symmetric.} 
\label{fig7}
\end{figure}

\begin{figure}
\centerline{
\hbox {
        \epsfxsize=11.8cm
        \epsfbox{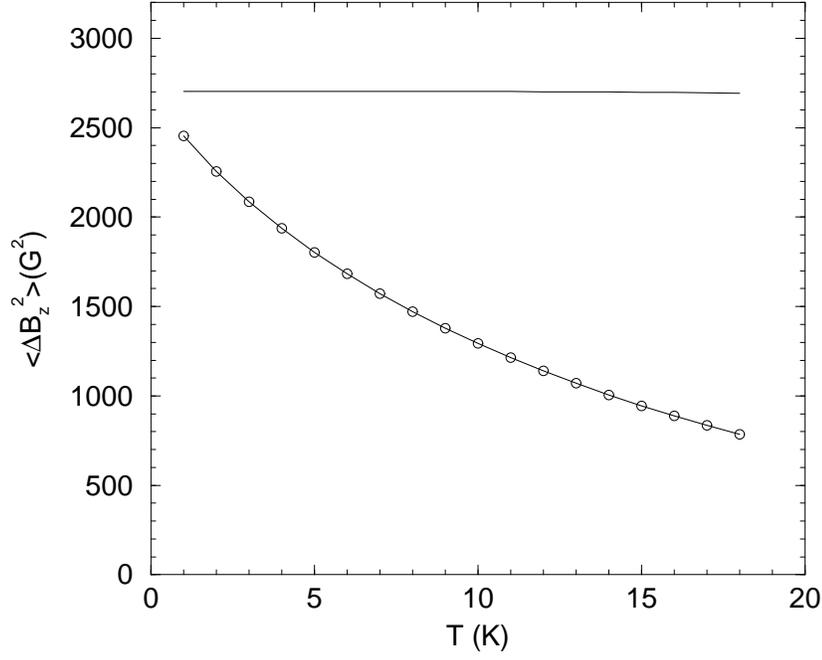}
        \vspace*{0.5cm}
       }
       }     
\caption{The second moment $[\Delta B^2] $ 
of the magnetic field distribution
$n(B)\/$ versus temperature $T$ for $B_0$ = 0.3 T, obtained using
the results of density functional theory for the Fourier
components of the density field and a Debye-Waller approximation
for the low-temperature regime. For comparison, the results for
a perfect Abrikosov lattice are also shown (solid line). Corrections
due to a finite core size (see text) are included. Note the
curvature of the moment as a function of $T\/$, which is
opposite to that obtained for a perfect Abrikosov lattice 
assuming a two-fluid form for the temperature dependence of
$\lambda$.} 
\label{fig8}
\end{figure}

\begin{figure}
\centerline{
\hbox {
        \epsfxsize=11.8cm
        \epsfbox{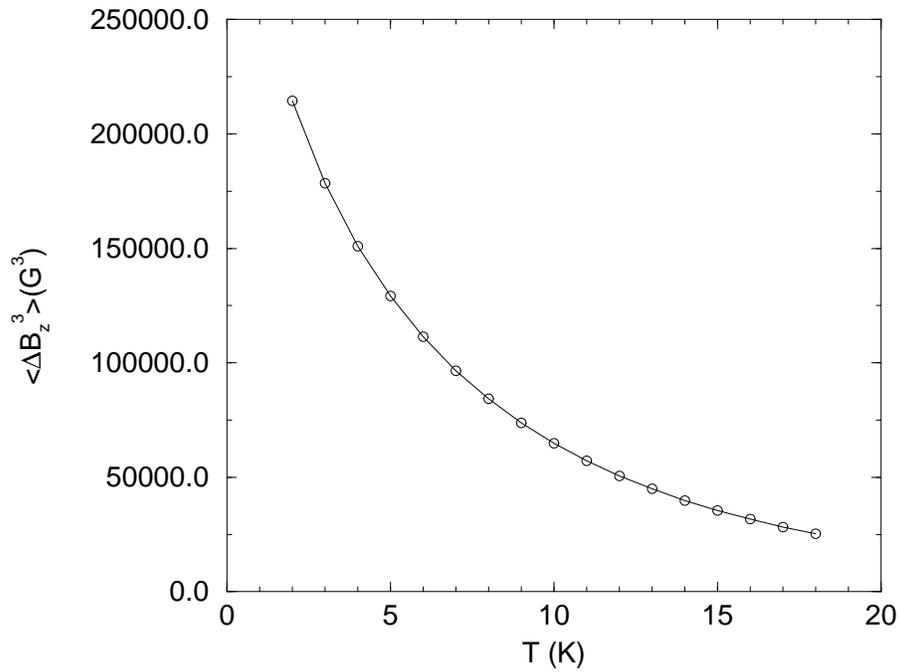}
        \vspace*{0.5cm}
       }
       }     
\caption{The third moment $[\Delta B^3]$ 
of the magnetic field distribution
$n(B)\/$ versus temperature $T$ for $B_0$ = 0.3 T obtained using
the results of density functional theory for the Fourier
components of the density field at the freezing transition
and a Debye-Waller approximation for the low-temperature regime. 
Corrections due to a finite core size (see text) are included.}
\label{fig9}
\end{figure}

\end{document}